\documentclass[prb,showpacs,preprintnumbers,preprint]{revtex4}
\usepackage{graphicx}
\begin{document}
\def\rhov{{\mbox{\boldmath{$\rho$}}}}
\def\tauv{{\mbox{\boldmath{$\tau$}}}}
\def\sigmav{{\mbox{\boldmath{$\sigma$}}}}
\def\xiv{{\mbox{\boldmath{$\xi$}}}}
\def\oh{{\scriptsize 1 \over \scriptsize 2}}
\def\of{{\scriptsize 1 \over \scriptsize 4}}
\def\tf{{\scriptsize 3 \over \scriptsize 4}}
\draft
\title{Ferroelectricity in Incommensurate Magnets}
\date{\today}
\author{A. B. Harris$^{a}$ and G. Lawes$^{b}$}

\address{a) Department of Physics and Astronomy, University
of Pennsylvania, Philadelphia, PA, 19104}
\address{b) Department of Physics and Astronomy, Wayne State University,
Detroit, MI, 48202}
\begin{abstract}
We review the phenomenology of coupled magnetic and electric order parameters
for systems in which ferroelectric and incommensurate magnetic
order occur simultaneously.  We discuss the
role that such materials might play in fabricating novel magnetoelectric
devices.  Then we briefly review the
mean-field description of ferroelectricity and modulated magnetic
ordering as a preliminary to analyzing the
symmetry of the interaction between the spontaneous polarization
and the order parameters describing long-range modulated magnetic
ordering.  As illustration we show how this formulation
provides a phenomenological explanation for the observed
phase transitions in Ni$_3$V$_2$O$_8$ and TbMnO$_3$ in which
ferroelectric and magnetic order parameters simultaneously
become nonzero at a single phase transition.  In addition,
this approach explains the fact that the spontaneous polarization
only appears along a specific crystallographic direction.  We analyze the
symmetry of the strain dependence of the exchange tensor and show that
it is consistent with the macroscopic symmetry analysis.  
We conclude with a brief discussion of how our approach might be relevant in
understanding other systems with coupled magnetic and ferroelectric
order, and more importantly, how these principles relate to the search
for materials with larger magnetoelectric couplings at room temperature.
\end{abstract}
\maketitle

\section {Introduction}

The interactions between long-range magnetic order and long-range
ferroelectric order have been studied in depth since the first
experimental confirmation of the magnetoelectric effect in the late
1950s.\cite{DZYALOSHINSKII,ASTROV,RADO} We
note the existence of several reviews\cite{REV1,REV2,REV3} and
monographs\cite{BIRSS} which give a general overview of the subject.

Of particular interest for this review are those materials which exhibit
a combined magnetic and ferroelectric transition.  Perhaps
the best known of these is Ni-I boracite (Ni$_3$B$_7$O$_{13}$I) which
shows coupled ferromagnetic, ferroelectric, and ferroelastric properties
at a single phase transition at $T=61.5$ K.\cite{ascher,toledano}  The
multiferroic behavior in this boracite arises from the fact that the
magnetic transition is connected to a structural distortion,
which in turn allows the development of ferroelectric order.\cite{toledano}
This transition can be understood in terms of a phenomenological Landau
theory which couples the ferromagnetic, ferroelectric, and ferroelastic
order parameters to a primary antiferromagnetic order parameter.\cite{toledano}
The strong coupling between magnetic and ferroelectric order parameters
in systems having a simultaneous phase transition is demonstrated by
the observation that in Ni$_3$B$_7$O$_{13}$I it is possible to reverse
the direction of the spontaneous polarization by applying an external
magnetic field perpendicular to the direction of magnetization.\cite{ascher}

Cr$_2$BeO$_4$ also develops magnetic and ferroelectric order at a single
phase transition.\cite{newnham} Below T=28 K, Cr$_2$BeO$_4$ orders
antiferromagnetically into a state with spiral spin structure, and this
antiferromagnetic state shows an extremely
small spontaneous polarization (approximately one million
times smaller than that of BaTiO$_3$).  The coupling
between magnetic and ferroelectric order is expressed by a model
proposing a mechanism in which the electric polarization is induced
solely by the antiferromagnetic order.\cite{stefanovskii}  A
similar model for magnetically-induced ferroelectric
order will be discussed in detail in the following sections. 

While the magnetic and ferroelectric transition temperatures for BaMnF$_4$
are widely separated, this system is useful in illustrating the importance
of symmetry considerations in determining magnetoelectric properties.
Pyroelectric BaMnF$_4$ orders antiferromagnetically when cooled below
$T=26$ K, and there is a dielectric anomaly at this magnetic transition
temperature.\cite{samara}  This decrease in dielectric constant below $T_N$
varies like the square of the sublattice magnetization, and clearly
indicates a coupling between the magnetic and ferroelectric properties of
the sample.  This interaction between magnetic and ferroelectric order
is attributed to a magnetoelectric coupling which causes a polarization
induced spin canting.\cite{foxA}  Substituting 1\%Co for Mn changes the
magnetic symmetry group of the compound to one which precludes this
magnetoelectric coupling,\cite{foxB} and in
turn eliminates the dielectric anomaly at the magnetic ordering
temperature.  As this system illustrates, in
order to understand magnetoelectric couplings in
multiferroic systems it is crucial to have complete information about
the magnetic and structural symmetries of the system.

Until quite recently, the theoretical and experimental studies have
focussed on ferroelectricity in systems with simple ferromagnetic or
antiferromagnetic order\cite{foxB,toledano} (with studies on Cr$_2$BeO$_4$
being the notable exception).  
These systems are tractable from a theoretical standpoint, and allow
a comparison to be made between experimental results and straightforward
models based on magnetic space groups.  However, limiting the scope
of investigation to systems with ferromagnetic or antiferromagnetic
order neglects a large class of materials which have more complex magnetic
structures.  Here we will not consider systems (several of which are listed
in Table I of Ref. \onlinecite{REV1})
which are ferroelectric at high temperature
and then have a lower temperature phase transition at which magnetic
ordering takes place.\cite{LRO} Instead, in this brief review article we
will focus on the more recent studies in which ferroelectricity appears
simultaneously (in a single combined phase transition) with long-range
sinusoidally modulated magnetic order,\cite{KIM,HUR} which we will refer
to generically as ``incommensurate"\cite{INCOM}  magnetic order.
Accordingly, we will briefly summarize the experimental situation for
the systems TbMnO$_3$ (TMO)\cite{HUR,TMO} and Ni$_3$V$_2$O$_8$
(NVO).\cite{PRL,RAPID,NVO1}  Then we will describe in detail the symmetry
analysis developed in Refs. \onlinecite{RAPID},
\onlinecite{NVO1}, and \onlinecite{NVO2}
to understand the phenomenology of these systems.  We believe that
this theoretical approach is simple
enough that it can easily be applied to the ever increasing number of
systems like NVO or TMO in which ferroelectricity is induced by
incommensurate magnetic long-range order. 

\begin{figure}
\begin{center}
\includegraphics[width=15cm]{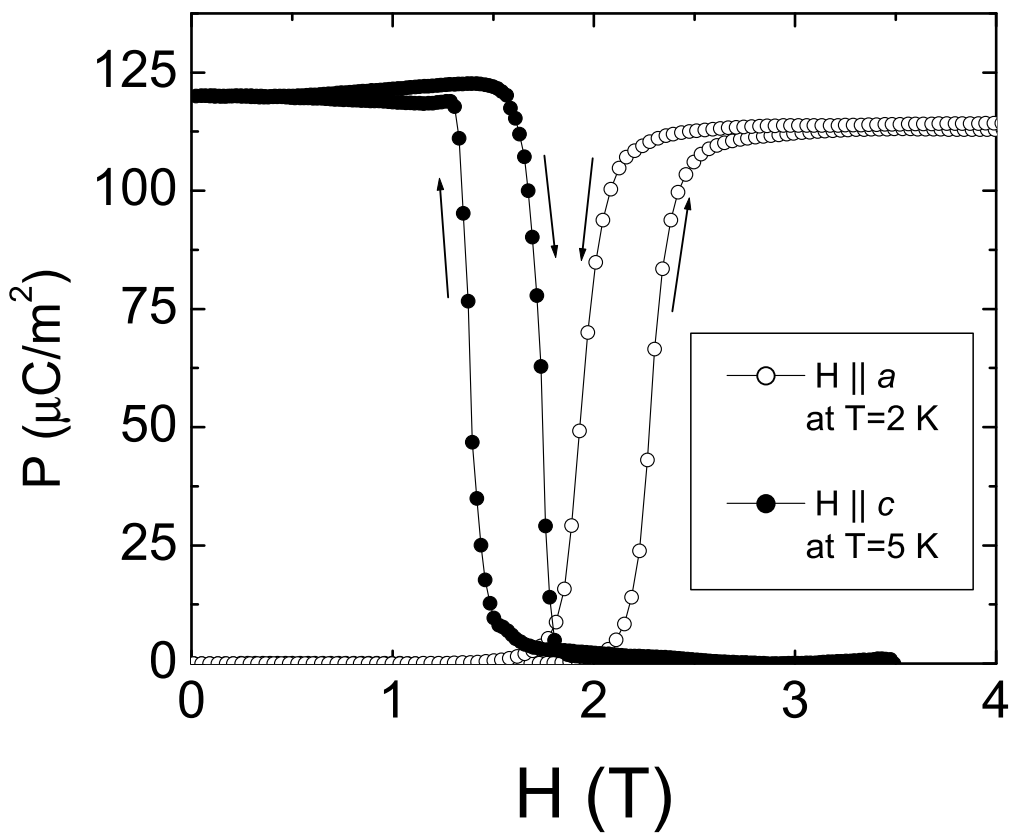}
\caption{Adapted from Ref. \protect{\onlinecite{RAPID}}.
Spontaneous ({\it i. e.} in zero applied electric field)
polarization ${\bf P}$ versus applied uniform magnetic field  ${\bf H}$
applied along different crystallographic directions at
selected temperatures for NVO. The arrows indicate the directions of
increasing and decreasing magnetic field.}
\label{GL3}
\end{center}
\end{figure}

To illustrate this phenomenon, we show, in Fig. \ref{GL3} some
intriguing data from Ref. \onlinecite{RAPID} showing that the
spontaneous polarization {\bf P} depends strongly on the
applied magnetic field {\bf H}.
At first glance this data seems to have no obvious explanation.
However, when viewed in combination with the magnetic phase diagram
(see Fig. \ref{PHASE}, below)
we will see that this data indicates that the spontaneous polarization
is nonzero only in the magnetic phase we will call the ``low temperature
incommensurate phase."  The hysteresis is a consequence of passing
through a first order phase boundary between this phase and an
antiferromagnetic phase in which a spontaneous polarization is
not allowed.  Thus the dramatic dependence of polarization on 
magnetic field has a simple explanation: ferroelectric order
appears only in one specific magnetic phase whose existence depends
in the value of the magnetic field.
This strong coupling between magnetic and ferroelectric order is
potentially important for device applications,
as we will discuss in the following section.  From a basic physics
standpoint, these systems which exhibit a coupling between
the ferroelectric moment (a polar vector) and the magnetic moment
(an axial vector) are very interesting.  (As we will see,
such systems have order parameters whose response to both electric and
magnetic fields becomes large especially near a phase transition.)
A complete understanding of this coupling from a microscopic theory is
not yet available.  Here we will show that the Landau expansion
explains the observed phenomenology of this interaction and that
these results follow from the
microscopic symmetry of the strain dependence of the exchange tensor.
This explanation will serve as a guide to constructing a fully
microscopic theory of magnetoelectric coupling.

Briefly, this review is organized as follows.  In Sec. II we discuss
some general types of applications in which the magnetoelectric
coupling may be exploited to develop new types of devices.  It
should be emphasized that these applications are speculative,
and are intended to illustrate the types of new devices that
could be developed using these new materials.
In Sec. III we review the Landau description of ferroelectricity.
In Sec. IV we give a simplified theoretical analysis of
incommensurate magnetic ordering and in Sec. V
we discuss how Landau theory leads to a symmetry-based description
of incommensurate magnetic ordering.  It is our aim to demystify the
use of representation theory for the determination of  magnetic structure
by diffraction techniques.  Understanding these incommensurate
magnetic structures is crucial to developing a model for the
coupling between magnetic and ferroelectric order in these systems.
In Sec. VI we use the results of Sec. V to analyze how symmetry
restricts the form of the coupling between electric and
magnetic order parameters and thereby explain the simultaneous
appearance of these two kinds of order parameters in a single phase
transition.  The construction of this interaction is greatly
simplified by the fact that it involves an expansion in powers of
the order parameters relative to the paramagnetic paraelectric
phase.  Thus the interactions have to satisfy the invariances of
the disordered paramagnetic/paraelectric phase\cite{PSYM1,PSYM2}
and we do not
need to broach the more complicated question of analyzing the symmetry
of interaction within an ordered phase.  In Sec. VII we analyze the
symmetry of the strain dependence of the exchange tensor and show
that it leads to results identical to those of Landau theory.
Finally in Sec. VIII we summarize the main points of this review and
speculate on some future directions of research.  We will discuss
how our results on ferroelectric order in incommensurate magnets
may offer guidance in searching for new magnetoelectric materials.

\section {Device Applications}

The development of  devices incorporating both charge and spin degrees of 
freedom, often referred to as {\it spintronics}, has already led to significant 
technological breakthroughs.\cite{SPINTRONICS}  Magnetic sensors based on
giant magnetoresistance (GMR) are widely used as the read heads in modern 
hard drives, and magnetic random access memory also relies strongly on couplings
between charge and spin. 
Additionally, there are a wide range of proposals 
for devices based on controlling the spin degree of freedom in 
ferromagnetic semiconductors, including spin valves and qubits for quantum 
computing.  Much of the research on materials in which charge and
spin are coupled have focussed on metallic and semiconducting systems.
However, dielectric materials exhibiting couplings between electric
polarization and magnetization may also play an important role in
developing the next generation of spintronic devices.

Magnetoelectrics are systems in which either applying an external magnetic
field produces an electric polarization or applying an external
electric field produces a magnetization.
This type of coupling between charge and spin was postulated by Pierre
Curie at the end of the 19th century,\cite{CURIE} but not observed
experimentally until the late 1950s.\cite{ASTROV,RADO}  Materials in
which two or more of ferroelectric, ferromagnetic, and ferroelastic
order coexist are referred to as multiferroics.  This strict definition
of multiferroics is often relaxed to include systems which exhibit
combinations of
any type of long range magnetic, ferroelectric, or ferroelastic order.
This review will concentrate specifically on magnetoelectric
multiferroics, where the coexistence of long range magnetic and
long range dielectric order leads to a pronounced couplings between
the charge and spin degrees of freedom in these systems.

We consider two classes of devices based on magnetoelectric 
multiferroics.  The first class of devices depend on the 
magnetoelectric effect---the induction of a magnetization (polarization) 
by an applied electric (magnetic) field.  Using the magnetoelectric 
effect, it is possible to design a range of devices from sensors to 
transducers to actuators, coupling magnetic and electric properties.  The 
second class of devices exploits the fact that these materials have 
simultaneously appearing long 
range magnetic and ferroelectric order. The underlying assumption is
that multiferroics exhibit both charge and spin ordering, and due to
the coupling between the two, both magnetic and ferroelectric
order will be strongly affected by either magnetic or electric fields.  Strictly
speaking, only magnetic field control of the electric polarization has
been demonstrated for the multiferroic materials with incommensurate
magnetic structures discussed in this review, but magnetic phase control
by an electric field has been demonstrated in other multiferroic
materials.\cite{PHASECONTROL}  This coupling between long range electric
and magnetic order leads to new functionalities which can be exploited 
for designing new types of spintronic devices.

The investigation of magnetoelectric devices is an active area of 
research.  Prototype devices fabricated using 
piezoelectric-magnetostrictive composite materials to produce 
magnetoelectric  coupling have already been tested,\cite{BAYRASHEV,GOPAL}
and there are a range 
of proposals for other magnetoelectric devices.  These include utilizing 
magnetoelectric materials as the pinning layer in GMR devices,\cite{BINEK}
for low frequency wireless power applications,\cite{BAYRASHEV} and for
developing tunable dielectric materials.\cite{TAKAGI}  
One key feature of magnetoelectric materials is that they allow 
the design of devices controlled magnetically or electrically, as desired.
Controlling the magnetic properties of materials using an {\it electric} field
offers significant benefits in designing new devices.  Using current-based
methods to switch magnetic devices is relatively slow, and power-intensive.
Voltage control of the magnetic properties is expected to offer significantly
faster switching (thin film ferroelectrics can show switching times of less
than ~200 ps \cite{rameshFE}) in a low-power device.
Magnetoelectric materials offer
the potential for fabricating highly tunable, fast switching, low-loss/low-power
devices having very small form factors, which would be suitable for a wide
range of commercial and industrial applications.

The materials property most relevant in determining the suitability of a 
compound for applications in magnetoelectric devices is the magnitude of 
the magnetoelectric susceptibility, $\chi_{ME}$.  For homogeneous materials,
$\chi_{ME}$ satisfies the bound,
\begin{eqnarray}
\left( \chi_{ME}^2\leq \chi_E \chi_M \right)
\end{eqnarray}

\noindent
where $\chi_E$ and $\chi_M$ are the electric and magnetic susceptibilities 
of the system respectively.\cite{REV1}  Therefore, in order to
maximize the magnitude 
of the magnetoelectric coupling, one should attempt to maximize the 
magnitudes of {\it both} $\chi_M$ and $\chi_E$.  Since ferroelectrics 
typically have large values of $\chi_E$ and ferromagnets typically have 
large values of $\chi_M$, multiferroics are expected to have large values 
of $\chi_{ME}$.  Furthermore, since susceptibilities are largest at the 
ordering transition, systems developing magnetic and ferroelectric order at 
the same temperature should show exceptionally large magnetoelectric 
couplings.  This has been confirmed for the intrinsic multiferroic 
Ba$_{0.5}$Sr$_{1.5}$Zn$_2$Fe$_{12}$O$_{22}$, which has
the largest magnetoelectric coefficient of any single-phase material 
identified to date.\cite{KIMURAFERRITE}  Understanding the microscopic
origins of the magnetoelectric coupling in these multiferroic systems
will have important 
ramifications for developing novel magnetoelectronic devices.

Beyond simply exhibiting very large magnetoelectric couplings, 
intrinsic multiferroics also have both long range magnetic order
and long range ferroelectric order.  The coupling between
magnetization and polarization offer new possibilities for designing
devices.
The ability to control the magnetic or ferroelectric state of a system
using either a magnetic field or an electric field would offer the
ability to develop multifunctional memory elements, for example,
ferroelectric memory which can be written to 
using magnetic fields.  We will discuss two proposals for new
technologies which explicitly utilize the ferroelectric and magnetic
characters of magnetoelectric multiferroics.  It should be emphasized
that this discussion is meant only to illustrate some of the potential
applications arising from the  incorporation of multiferroic materials
into new devices.  More investigation on the specific properties of
these multiferroics is required before proof-of-principle devices
could be designed based on these speculations. 

As the bit density of modern hard drives increases, the characteristic size
of the magnetic structures used to store the information is decreasing.
As the physical size of the bit is reduced, the
anisotropy energy decreases, and the magnetic moment can begin to
thermally fluctuate.  Controlling these thermal fluctuations is
necessary to ensure the long-term stability of stored information in
ultra-dense magnetic recording material.  For long term magnetic 
storage (5+ years), the ratio of the energy barrier against these
thermal fluctuations to $k_BT$ should be large, roughly ~50.  In
current devices, this is often accomplished by using materials with very
large magnetic anisotropy energies or by exploiting the anisotropic
difference between FM and AFM layers.  One possible application for
multiferroic materials is to use the coupling between ferroelectric and
magnetic order in these systems to stabilize the magnetic moment against
thermal fluctuations in nanoscale magnets.

In many magnetoelectric multiferroics there is a strong coupling between
the ferroelectric and magnetic order parameters.  In such systems, fixing
the polarization (magnetization) direction will fix the axis of the
magnetization (polarization).   This coupling 
is observed in measurements showing
that  the sign of the magnetically induced polarization is independent of
the sign of the applied magnetic field, although the development of
ferroelectric order depends strongly on the magnetic field axis.
In such multiferroics, fixing the electric polarization would also fix
the magnetization axis.  This ferroelectrically induced magnetic anisotropy
would inhibit thermally activated switching of the magnetic moments by
significantly increasing the magnitude of the energy barrier to
magnetization reversal.  This could be accomplished, for example, by
assembling multiferroic nanoparticles on a ferroelectric substrate.  In
this geometry, the very large ferroelectric anisotropy energy
would provide a tunable barrier against thermal fluctuations of the
{\it magnetic} moment as well.

\begin{figure}
\begin{center}
\includegraphics[width=15cm]{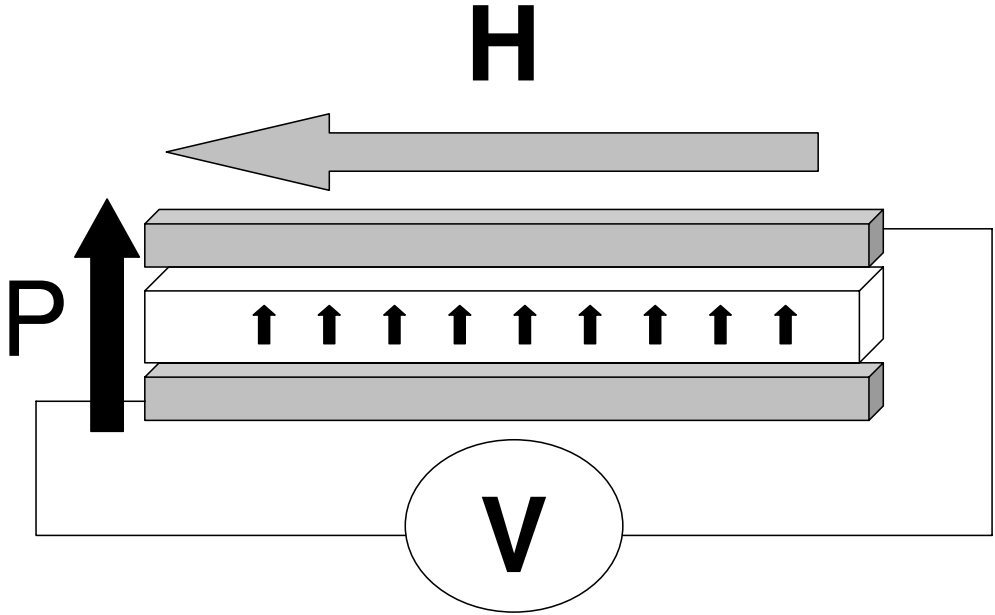}
\caption{Schematic illustration of a device to measure magnetic fields by
using the induced polarization.  The middle layer (white) is a multiferroic 
material with strong magnetoelectric coup0lings and the outer layers (gray) are
ferromagnetic metals. In this example the polarization is induced in a
direction perpendicular to the applied magnetic field.}
\label{NASAFIG}
\end{center}
\end{figure}

Multiferroics may also have important applications in developing
magnetic field sensors.  There are a range of proposals for incorporating
magnetoelectric materials in exceptionally sensitive magnetic field detectors.  
Even relatively small external magnetic fields will
produce a voltage change in materials with very large magnetoelectric
couplings.  Since it is often better to measure small voltages 
at zero applied current rather than small magnetizations or small changes in
resistivity, magnetoelectric materials offer the potential for developing
greatly improved magnetic field sensors.  Because multiferroics exhibiting 
simultaneous magnetic and ferroelectric transitions offer exceptionally large
magnetoelectric couplings, these materials are particularly interesting in
the context of improved sensors.  Figure \ref{NASAFIG} shows a schematic
for such a device.\cite{NASA}  The magnetization produces 
a spontaneous polarization directed perpendicular to the plane of the sensor.
This magnetically induced voltage can be measured to a high degree
of accuracy, either directly, or by measuring the dielectric
response of the compound.  This device could also be configured
to extract energy from an {\it alternating} magnetic field---the
magnetically induced alternating voltage could be used as a supply for
very low power applications.\cite{BAYRASHEV}

\begin{figure}
\begin{center}
\includegraphics[width=12cm]{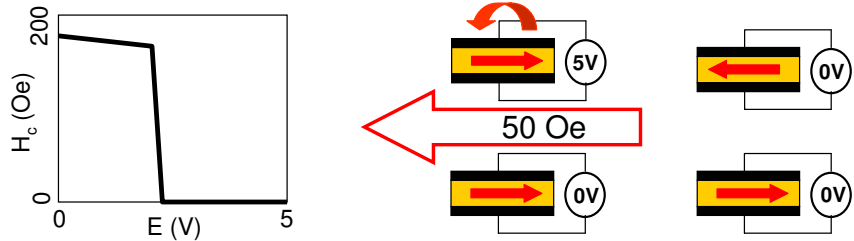}
\caption{Voltage-assisted MRAM. Left: Schematic diagram showing how the
coercive magnetic field is expected to vary with voltage.  Right:
With the magnetization of the bits originally directed to the right, a
small (50 Oe) writing field is applied.  At zero voltage, this is smaller than
the coercive field, so that the magnetization does not switch.  However,
applying 5V across the multiferroic layer reduces the coercive field
to almost zero, so that this small magnetic field is sufficient to
reverse the direction of magnetization.}
\label{MRAMFIG}
\end{center}
\end{figure}

Beyond simply being used as a passive magnetic field sensor, the device
illustrated in Fig. \ref{NASAFIG} could also be configured as a voltage
biased magnetic memory element.  One of the difficulties facing
current magnetic random access memory (MRAM) devices lies in producing
sufficiently strong magnetic fields to cause a moment reversal in the
memory element, but also sufficiently localized to affect only one
specific element.  While identifying multiferroic materials in which
applying a voltage could reverse the direction of the magnetization
would certainly be beneficial for developing MRAM devices, a more
modest type of voltage-assisted magnetization reversal could also
be significant.  As will be discussed in the following section,
ferroelectric order can
be promoted or suppressed by the application of an external magnetic
field in many multiferroic materials.  We expect that in these
materials, applying an electric field could then suppress or
promote magnetic ordering.  In such a system, the coercivity of
the magnetic memory element could be tuned by applying an electric
field.  Consider the multiferroic memory element in a ferromagnetic
state, which can be suppressed by applying a sufficiently large voltage.
In the absence of an electric field, the coercivity of the memory
element is large, so the magnetization is unaffected by stray
magnetic fields.  In order to reverse the magnetization direction,
a bias voltage is applied to the multiferroic, bringing the system
closer to the magnetic transition, reducing the magnitude of the
coercive field.  In this state, the magnetization can be reversed by
a relatively small external magnetic field, smaller than the coercive
field of the unbiased multiferroic.  When the voltage is removed, the
new magnetization will be stable.  This type of voltage-assisted
magnetization reversal could be used to produce arrays of magnetic
memory elements which could be switched by the same external magnetic
field.  Only those elements which have a bias voltage applied will
have a sufficiently small coercivity to be switched by the magnetic
field.  This technique may offer advantages over transitional MRAM
devices, such as a smaller sensitivity to stray
fields (allowing higher bit density) and potentially faster switching
times.  This is schematically illustrated in Fig. \ref{MRAMFIG}.

\section{FERROELECTRICITY}

We start by making a few observations concerning the symmetry properties
of ferroelectric systems for which magnetic ordering plays no role.
In the most common scenario, ferroelectrics
exhibit a high-temperature phase having spatial inversion symmetry
which prevents the existence of a vector order parameter.  Then,
as the temperature is reduced through a critical value, $T_F$,
a lattice instability develops in which inversion symmetry is broken
cooperatively via a continuous phase transition
at which a spontaneous polarization appears. Within a Landau theory
this transition is described by a free energy of the form
\begin{eqnarray}
F &=& \oh \chi_E^{-1} {\bf P}^2 + {\cal O} ({\bf P})^4
= \oh a (T-T_F) {\bf P}^2 + {\cal O} ({\bf P})^4 \ .
\label{FFE} \end{eqnarray}
At the transition the fact that the quadratic term in ${\bf P}$ becomes
unstable (negative) reflects the divergence in the electric susceptibility
at the ferroelectric transition.  This instability is sometimes traced
to a soft phonon, but whatever the mechanism, the appearance of
ferroelectricity represents a broken symmetry.  Conversely, as will
become relevant in the following, ferroelectricity can only occur if the
symmetry is broken to permit the ordering of the polarization vector.
We will use this criterion to determine which types of magnetic order
can possibly induce ferroelectric order.  If one takes the quartic
terms in Eq. (\ref{FFE}) to be of the form $a[{\bf P}^2]^2$ (with
$a>0$ for stability), then minimization of $F$ with respect to ${\bf P}$
shows that for $T<T_F$ one has $P \sim [2a(T_F-T)]^{1/2}$, which
is expected to hold as long as $T_F-T$ is not so large that sixth
and higher order terms in $F$ are important.  Mean field theory ignores
spatial correlations which lead to modifications of critical exponents,
but the scope of this review does not permit consideration of such 
corrections.\cite{PFEUTY}

As the temperature is further lowered it is possible for this
ferroelectric system to develop long-range magnetic order.\cite{LRO}  In
this case, one does not expect significant interaction between electrical
and magnetic properties because the two phenomena are essentially independent
of one another.  In these systems, the spontaneous polarization
will depend only weakly on the applied magnetic field.  In this
scenario, it is well known\cite{FEREF} that one can expect anomalies in
the dielectric response of the system when the ferroelectric develops
(independently) long-range magnetic order.
This review is {\it not} concerned with  such an ``accidental''
superposition of electric and magnetic properties. Instead  
we focus our attention on the situation when the appearance of
long-range magnetic order induces ferroelectricity.  Furthermore,
we will consider an interesting subclass in which the long-range
magnetic order is modulated with an apparently incommensurate
wavevector.  We will develop a Landau theory for this
combined phase transition in which the fact that the wavevector
does not have high-symmetry (and is thus neither ferromagnetic
or antiferromagnetic) is crucial to our analysis.  Thus
the development here can not be obtained by a trivial extension
of theories applicable to ferro- or antiferromagnetic ferroelectrics.
A simplifying feature of this formulation is that it is based
on an expansion of the free energy in powers of the various
order parameters relative to the {\it paramagnetic} phase.
Accordingly, each term in this expansion has to have the full
symmetry of the disordered phase.\cite{PSYM1,PSYM2}  In contrast, it
is less straightforward to analyze whether or not the symmetry of a
magnetically ordered phase permits an induced ferroelectric order.
Also, the Landau formulation correctly predicts
which components of the spontaneous polarization vector are
induced by the magnetic ordering.  In addition, the Landau expansion
indicates that the spontaneous polarization is, crudely speaking,
proportional to the emerging magnetic order parameter.

\section{TOY MODELS FOR INCOMMENSURATE MAGNETISM}

\subsection{Review of Mean Field Theory}

In this section we review the description and phenomenology of
incommensurate magnets, because the characterization of their
symmetry is essential to understanding the coupling between
magnetic and electric long range order.

For the purposes of this review it suffices to consider the
description of incommensurate magnets within mean field theory.
For a system consisting of quantum spins of magnitude $S$ on each
site, we write the trial free energy is
\begin{eqnarray}
F \equiv U-TS = {\rm Tr} [ \rhov {\cal H} + kT \rhov  \ln \rhov ] \ ,
\label{FEQ} \end{eqnarray}
where ${\cal H}$ is the Hamiltonian, $T$ the temperature, $U$ the
internal energy, $S$ the entropy, and
the actual free energy is the minimum of $F$ with respect to the
choice of $\rhov$ subject to the conditions that $\rhov$ is
Hermitian with unit trace.  Within mean field theory we take
the density matrix to be the product of independent single
particle density matrices $\rhov (i)$ for each site $i$:
\begin{eqnarray}
\rhov \equiv \prod_i \rhov (i)\ .
\end{eqnarray}
This approximation corresponds to the intuitive idea that when correlations
between spins are neglected, each spin reacts to the mean field of its
neighbors.

In Eq. (\ref{FEQ}) the trace of $\rhov {\cal H}$ gives the internal
energy $U$ and that of $-k \rhov \ln \rhov$  gives the entropy $S$.
In the absence of anisotropy it suffices to set
\begin{eqnarray}
\rhov (i) &=& {1 \over 2S+1} [ {\cal I} + c \sigmav (i) \cdot {\bf S}_i ] \ ,
\end{eqnarray}
where ${\cal I}$ is the unit matrix of dimension $(2S+1)$, $c$ is a
constant of order unity, chosen to make Eq. (6) true,
and ${\bf S}_i$ is the vector spin  operator for site $i$
[Here ${\bf S}_i$ is a $(2S+1)$ dimensional matrix]. The free
energy is then minimized with respect to the trial parameters $\sigmav (i)$,
which physically are identified as the average spin vectors:
\begin{eqnarray}
\langle {\bf S}(i) \rangle \equiv {\rm Tr} [\rhov(i) {\bf S}(i)]
= \sigmav (i) \ .
\end{eqnarray}
Thus $\sigmav (i)$ is the vector {\it order parameter} at the $i$th
lattice site.  In this formulation the internal energy is quadratic in
the order parameter $\sigmav$, whereas the entropic term involves both
quadratic
and higher powers of the order parameter.  As we shall see, even without
explicit calculations much information can be inferred from the symmetry
of the trial free energy as a function of the order parameter(s).

As mentioned in the introduction, we will focus our attention on
systems which display incommensurately modulated magnetic long range
order.  We refer the reader to a comprehensive survey of such
systems by Nagamiya.\cite{NAG}  Here we give a simplified review. 
To characterize an incommensurate state we consider a toy model
consisting of a one dimensional system with isotropic antiferromagnetic
exchange interactions $J_1$ and $J_2$ between nearest and next-nearest
neighbors, respectively.  If $J_2$ is antiferromagnetic and large enough,
these two interactions compete and produce an incommensurate spin structure.
Thus we are led to consider the Hamiltonian
\begin{eqnarray}
{\cal H} = \sum_n {\bf S}_n \cdot [J_1 {\bf S}_{n+1} + J_2 {\bf S}_{n+2}] \ ,
\end{eqnarray} 
with $J_2>0$.  The corresponding trial free energy is
\begin{eqnarray}
F &=& \oh dkT \sum_i  \sigmav(i)^2 + \sum_{n=1,2} J_n \sigmav(i) \cdot
\sigmav (i+n) + {\cal O} (\sigmav^4) \ ,
\end{eqnarray}
where the entropic term is scaled by a constant of order unity, $d$.

\subsection{Wavevector Selection}

\begin{figure}[ht]
\vspace{1.4 in} \begin{center}
\includegraphics[width=9cm]{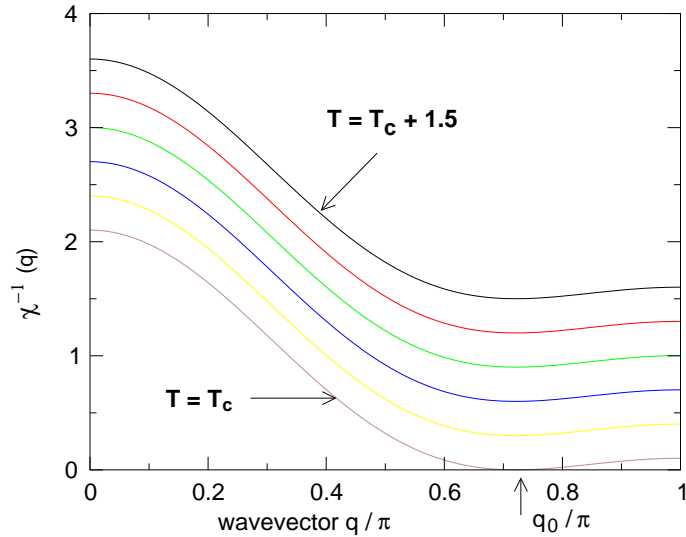}
\caption{$\chi(q)^{-1}$ at a sequence of temperatures, $T=T_c+0.3n$
for $n=0,1,2,3,4,5$ with $J_1=1.0$ and $J_1/J_2=2.56$ (as is
appropriate for NVO). Here $T_c= 0.711$.}
\label{CHIFIG}
\end{center}
\end{figure}

It is instructive to write the free energy per spin, $f$, in terms of
Fourier variables, $\sigmav (q) = (1/N) \sum_i e^{i qx_i} \sigmav(i)$,
where $N$ is the total number of spins as
\begin{eqnarray}
f &\equiv& F/N = \oh \sum_q \chi(q)^{-1} \sigmav(q) \cdot \sigmav(-q)
+ {\cal O} (\sigmav^4) \ ,
\end{eqnarray}
where $\chi(q)^{-1} = dkT + J_1 \cos (qa) + J_2 \cos (2qa)$ is
the wavevector-dependent susceptibility.
At high temperature (when $kT \gg |J_1|$ and $kT \gg |J_2|$), 
$\chi(q)^{-1}$ is positive for all $q$ and the free energy
is minimized by setting all the order parameters $\sigmav (q)$ to zero.
In Fig. \ref{CHIFIG}  we show $\chi(q)^{-1}$ as a function of wavevector
$q$ for a sequence of temperatures.  As the temperature is lowered through
a critical value $T_c$, $\chi(q)^{-1}$ becomes zero for
the wavevector $q\equiv q_0$ which minimizes $\chi(q)^{-1}$:
\begin{eqnarray}
\cos (q_0a) = - J_1/(4J_2) \ .
\label{QEQ} \end{eqnarray}

This determination of the value of $q_0$ is called {\it wavevector
selection}.  As the temperature is reduced through $T_c$ the
paramagnetic phase becomes unstable against the formation of
long range order at the selected wavevector $q_0$.  That is,
for $T<T_c$ the order parameter $\sigmav (q_0)$ assumes a nonzero value
determined by the (negative) quadratic terms in combination with the
(positive) terms of order $\sigmav^4$, so that
$|\sigmav (q_0)| \sim (T_c-T)^{1/2}$.
Once order develops at one wavevector, the terms of order $\sigmav^4$
prevent order developing at other wavevectors.
This scenario is realistic for a three dimensional system
(for which long-range order is not destroyed by thermal fluctuations).
The eigenvector associated with the eigenvalue of the quadratic form
which passes through zero is called the {\it critical eigenvector}.
The critical eigenvector contains the form factor of the ordering,
{\it i. e.} it completely describes the pattern of spin ordering within
a unit cell. In this simple model there is only one spin
per unit cell, so the eigenvector specifies the
direction {\it i. e.} the component which condenses.  (This
concept will be better illustrated when we consider real systems
which often have more than one magnetic site per unit cell.)
In the present case when there is no anisotropy, the spin structure
when $\sigmav(q)$ becomes nonzero for $q=q_0$ is
a modulated one in which the $x$-component of spin has a complex
amplitude, $A_x$, so that
\begin{eqnarray}
\sigma_x(i) &=& A_x e^{iq_0x_i} + A_x^* e^{-iq_0x_i}
\end{eqnarray}
and similarly for the other spin components.
If these complex amplitudes $A_\alpha$ all have the same phase,\cite{PHASE}
then the spin is linearly polarized with an amplitude which varies
sinusoidally with position.  If the complex amplitudes do not
have the same phase, then the spin structure will be a helix, a spiral,
or a fan, etc. 

\subsection{Effects of Anisotropy}

\begin{figure}
\begin{center}
\includegraphics[width=8cm]{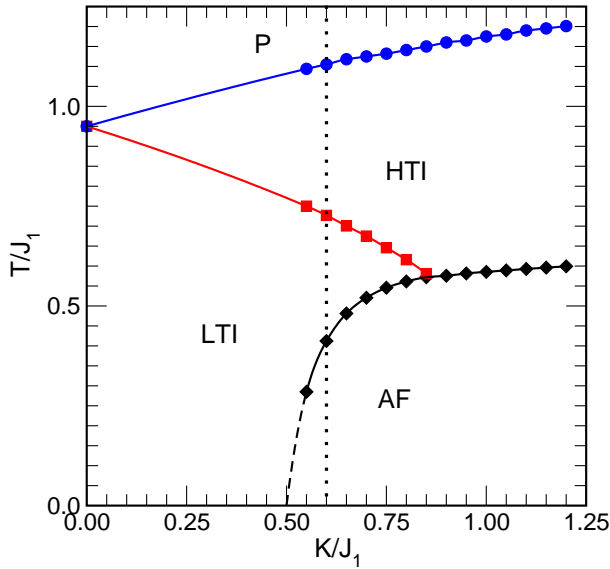}
\caption{Zero temperature phase diagram for the $J_1$-$J_2$ ($S=1$) model
with easy-axis anisotropy scaled by $K$.\protect{\cite{NVO1}}
The points represent
numerical implementation of mean-field theory except for $K=0$
where analytic results are used.   Here ``HTI" denotes a
longitudinally polarized incommensurate phase and ``LTI"
an elliptically polarized incommensurate phase.  In both
phases the modulation vector is given by Eq.  (\protect{\ref{QEQ}}).
``AF" denotes a two-sublattice collinear antiferromagnetic phase.
For large anisotropy this model reduces to the anisotropic nearest
next-nearest neighbor Ising (ANNNI) model.\protect{\cite{ANNNI}}
The dashed line is drawn for a value of the anisotropy energy
which reproduces the evolution of magnetic phases in NVO as a
function of $T$ for $H=0$.}
\label{TJHA}
\end{center}
\end{figure}

\begin{figure}
\begin{center}
\includegraphics[width=15cm]{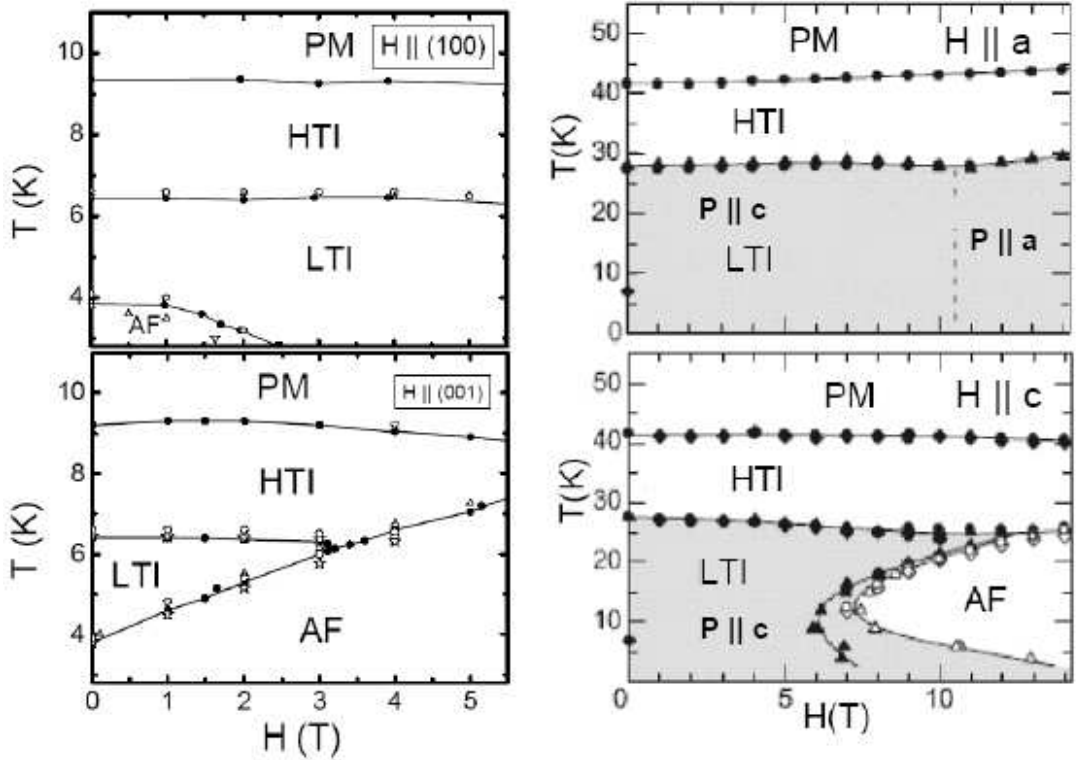}
\caption{Phase diagram for NVO (left) and TMO (right) as a function of
$H$ and $T$.  The HTI and LTI phases will be discussed in more detail below,
but correspond roughly to the scenario of our toy model.  The interpretation
of the $H$-dependence of the phase boundaries for NVO is given in Ref. 
\protect{\onlinecite{NVO1}}.  For NVO the C and C' phases are
commensurate antiferromagnetic phases with a weak ferromagnetic
moment.  The NVO phase diagram is adapted from Ref. \protect{\onlinecite{PRL}}
and the TMO phase diagram is adapted from Ref. \protect{\onlinecite{TKPRB}}.} 
\label{PHASE}
\end{center}
\end{figure}

This toy model will not accurately capture the behavior of real magnetic
systems because we have not yet included any anisotropy.
In the presence of single-ion easy axis anisotropy, the trial free energy
at quadratic order assumes the form
\begin{eqnarray}
f &=& \oh  \sum_q \chi(q)^{-1} \sigmav(q) \cdot \sigmav(-q) - K
\sum_q \sigmav_x (q) \sigmav_x(-q) + {\cal O} (\sigmav^4) \ ,
\end{eqnarray}
where $K$ is an anisotropy energy which favors alignment of spins
along the easy axis, here the $x$-axis, and $f$ denotes
the free energy per spin.  In this case, the instability
(at which long-range order first appears) is one in which
the spins are confined to the easy axis and have a sinusoidally varying
amplitude. This type of ordered phase will be referred to as
the high-temperature
incommensurate (HTI) phase and the associated critical temperature
will be denoted $T_{\rm HTI}$.  If the anisotropy is not too large,
then, as the temperature is further reduced, the fourth order
terms in the free energy (which we have so far neglected)
become important.  One effect of these terms can be visualized
as incorporating the constraint of ``fixed length."  In the
HTI phase the spins have a length which varies sinusoidally
with position.  However, in the ground state, we expect
each spin to have its maximum length $S$ but to be oriented
in a direction to optimize the energy.  Thus, in the extreme
limit of zero temperature, the constraint of fixed spin length
is fully enforced.  Although the constraint is
less fully realized at higher temperature, the qualitative effect
is clear: when the temperature is sufficiently reduced, one has a
continuous phase transition into a phase we refer to as the
low-temperature incommensurate (LTI) phase. In this phase the spins develop
transverse order (in addition to the preexisting longitudinal order
along the easy axis) to more nearly achieve fixed spin length.
If the easy axis anisotropy is small, the range of temperature over
which the HTI phase is stable is also small.  The phase diagram
of such a model as a function of anisotropy energy $K$ and
temperature is shown in  Fig. \ref{TJHA}.\cite{PRL,NVO1}  We will
mainly be concerned with the two incommensurate phases, the
longitudinally modulated HTI phase and the elliptically polarized
low-temperature
incommensurate LTI phase.  Although the details of the unit cell
complicate the picture, the phenomenology of the HTI and LTI phases
are usually roughly similar to that of the simplified case discussed here.
In Fig. \ref{PHASE} we show the experimentally determined phase diagrams
of NVO and TMO as a function of applied magnetic field $H$ and
temperature $T$, with the HTI and LTI phases labelled.

\subsection{Wavevector Locking}

\begin{figure}[ht]
\begin{center}
\includegraphics[width=16cm]{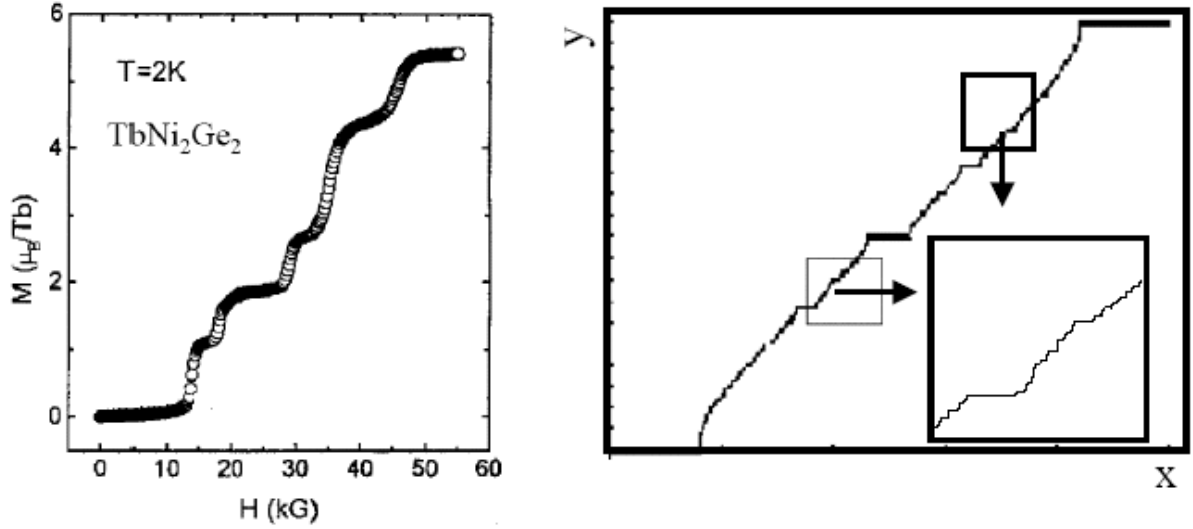}
\caption{The Devil's staircase (DS).  Left: an {\it incomplete} DS for
the dependence of the magnetization of TbNi$_2$Ge$_2$ as a function of
magnetic field.\protect{\cite{FIG7}} Right: Schematic example of
a {\it complete} DS function $Y=f(X)$. In a complete DS the function is
nonconstant on a set of measure zero.  For an incomplete DS the function
is nonconstant on a set of nonzero measure.}
\label{DEVIL}
\end{center}
\end{figure}

From Eq. (\ref{QEQ}) it would seem that the wavevector $q$ is a
continuous and smooth function of $J_2/J_1$.  Although our
toy model does not give any simple explanation for the observed
temperature dependence of $q$, a more complete analysis
[as in Refs. \onlinecite{NVO2} and \onlinecite{KAJI}]
leads to a small dependence on temperature
which, like the dependence on $J_2/J_1$, might be thought
to be smooth and continuous. However, there are terms which favor
commensurate values of $q$.  These terms in
the free energy must conserve wavevector, but only to within
a reciprocal lattice vector $G$ (which for our one-dimensional
toy model can assume the values $G_m = (2m \pi /a)$,
where $a$ is the nearest-neighbor separation).  Thus one has the
so-called Umklapp terms such as $\delta f = w \sigma (q)^4$, when
$4q = {\bf G}_m$.  More generally the Umklapp terms give a contribution
to the free energy of the form
\begin{eqnarray}
\delta f &=& \sum_{m,n} w_{m,n} \sigma(q)^{2n} \delta_{nq-2 m \pi /a} \ ,
\end{eqnarray}
where the coefficient $w_m$ is of order unity and $\delta_x$ is unity if
$x=0$ and is zero otherwise.  (Within the present formulation these terms
come from expanding ${\rm Tr} \rhov \ln \rhov$ to higher than quadratic 
order in the order parameter.) The effect of these Umklapp terms is
to cause the wavevector to ``lock" onto a commensurate value
$q=2(m/n)\pi /a$ as $J_2/J_1$ is varied.  Since $\sigma(q)$ is smaller
than one, especially near the ordering transition, these terms become
much less important 
as the integer denominator $n$ increases.  Thus the effect of the Umklapp
terms is that the variation of $q$ as a function of a control parameter
(such as the temperature) becomes a so-called Devil's staircase,
which may either be complete or (if $\sigma$ is small enough)
incomplete, as shown in Fig. \ref{DEVIL}.
In the systems we will discuss here, there is no clear
evidence of a Devil's staircase as a function of temperature.
Accordingly, we find it convenient to imagine that $q$ is incommensurate,
and does not get stuck on commensurate values by Umklapp terms.
Even if this is not strictly true, the difference in
properties between an incommensurate system and a commensurate
system with a large integer denominator $n$ is experimentally
irrelevant for the large values of $n$ ($n \sim 50$), for the
systems we will discuss.  Accordingly, we will refer to the systems
as ``incommensurate" even though this may not be strictly accurate.

In principle, the symmetry of real systems is usually such that
anisotropy also occurs in the exchange interaction, in which case
the trial free energy assumes the form
\begin{eqnarray}
f &=& \oh \sum_q \sum_{\alpha=x,y,z} \chi(q)_\alpha^{-1}
\sigmav_\alpha (q) \cdot \sigmav_\alpha (-q) + {\cal O} (\sigmav^4) \ ,
\end{eqnarray}
where $\chi(q)_\alpha^{-1} = dkT + J_{1\alpha} \cos (qa) + J_{2\alpha}
\cos(2qa)$.  If $J_{n\alpha}$ is isotropic (i.e. if it does not
depend on $\alpha$), then the wavevector selected for the ordering
of the $\alpha$ component of spin also will not depend on $\alpha$.
However, in principle $J_{n\alpha}$ depends weakly on $\alpha$, and
therefore the selected wavevector $q_0$ will also depend weakly on
$\alpha$ and the ordering will involve
$\sigma_x(q_x)$, $\sigma_y(q_y)$, and $\sigma_z(q_z)$.
Thus in the LTI phase it is possible that the two components of spin
might have slightly different wavevectors, which we denote
$q_{\rm HTI}$ and $q_{\rm LTI}$.  But as with the Umklapp contributions,
there will be quartic terms in the free energy (in this formulation
coming from the entropic terms) which favor locking the
two wavevectors to be equal.  These terms can be of the form
\begin{eqnarray}
f_{\rm lock} &=& a [\sigmav_{\rm HTI} (q_{\rm HTI})^2 
\sigmav_{\rm LTI} (-q_{\rm LTI})^2 +  \sigmav_{\rm HTI} (-q_{\rm HTI})^2
\sigmav_{\rm LTI} (q_{\rm LTI})^2 ] \delta_{q_{\rm HTI}, q_{\rm LTI}} \ ,
\end{eqnarray}
where $\sigma_{\rm HTI}e^{i\phi_{\rm HTI}}\equiv \sigmav_{\rm HTI}(q_{\rm HTI})$
($\sigma_{\rm LTI} e^{i\phi_{\rm LTI}}\equiv \sigmav_{\rm LTI}(q_{\rm LTI})$)
is an order parameter characterizing the appearance of the HTI (LTI) phase, and
for simplicity we have assumed that the constant $a$ is real-valued.
This interaction only satisfies wavevector conservation if the two
wavevectors are exactly equal.  If, in the absence of this term, the
two wavevectors are sufficiently close to one another, then this
locking energy will cause the wavevectors of the two order parameters
to be locked into equality with one another.  (In this case minimization
of $\delta f$ will also fix the relative phase $\phi_{\rm HTI}-\phi_{\rm LTI}$.)
Since exchange anisotropy is usually not large, the wavevectors associated
with different spin components are normally almost equal to one another.
In that case $f_{\rm lock}$ will be large enough to lock the HTI and LTI
wavevectors to a common value.  This ``locked'' scenario is
quite common and we assume it to be the case here.  Indeed for the systems
discussed below, the data suggests that the HTI and LTI order parameters
involve a single wavevector.  

\section{MAGNETIC SYMMETRY}

\subsection{Nontrivial Unit Cell}

\begin{center}
\begin{figure}[ht]
\includegraphics[width=8cm]{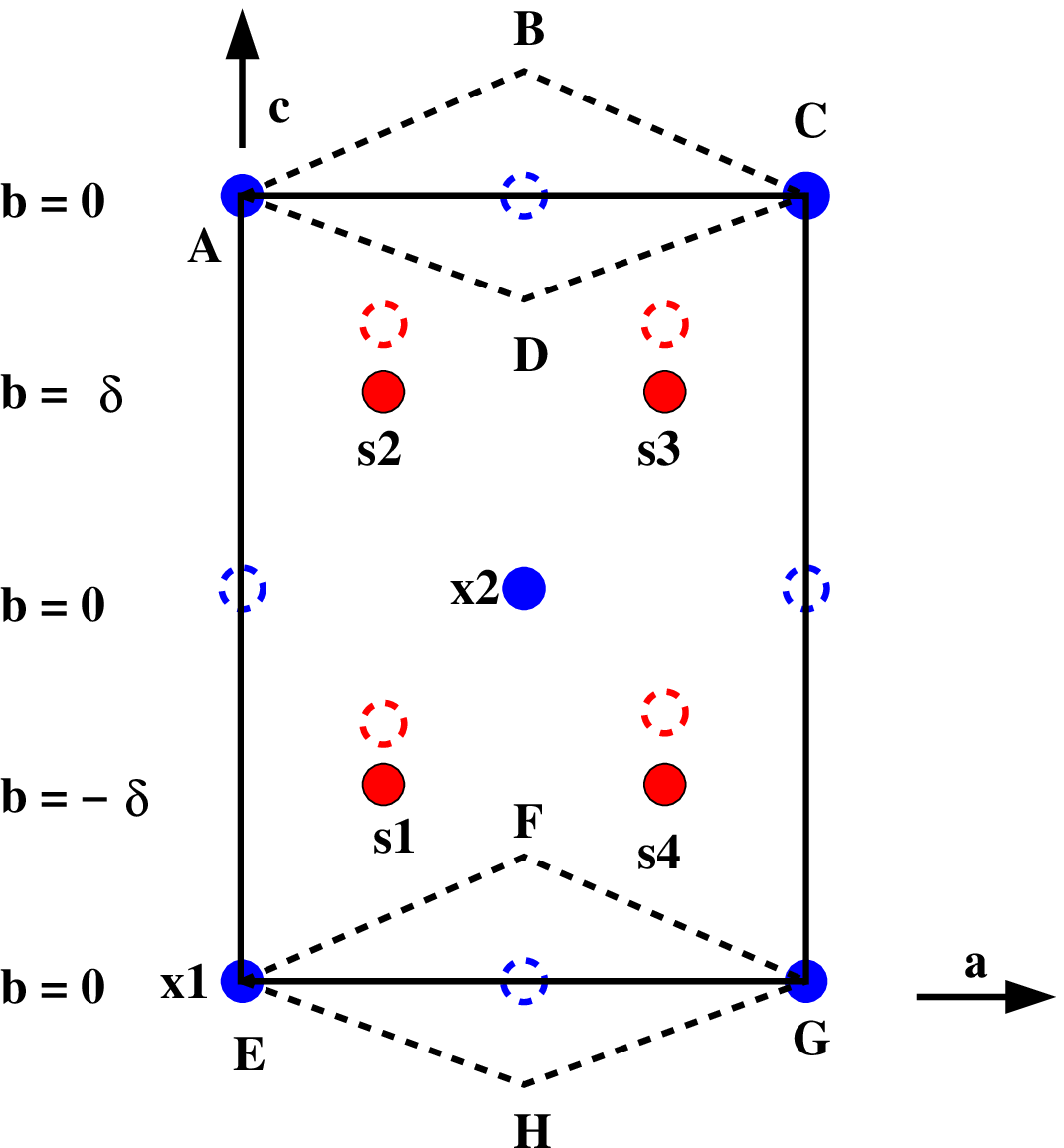} \hspace{0.3 in}
\includegraphics[height=8cm]{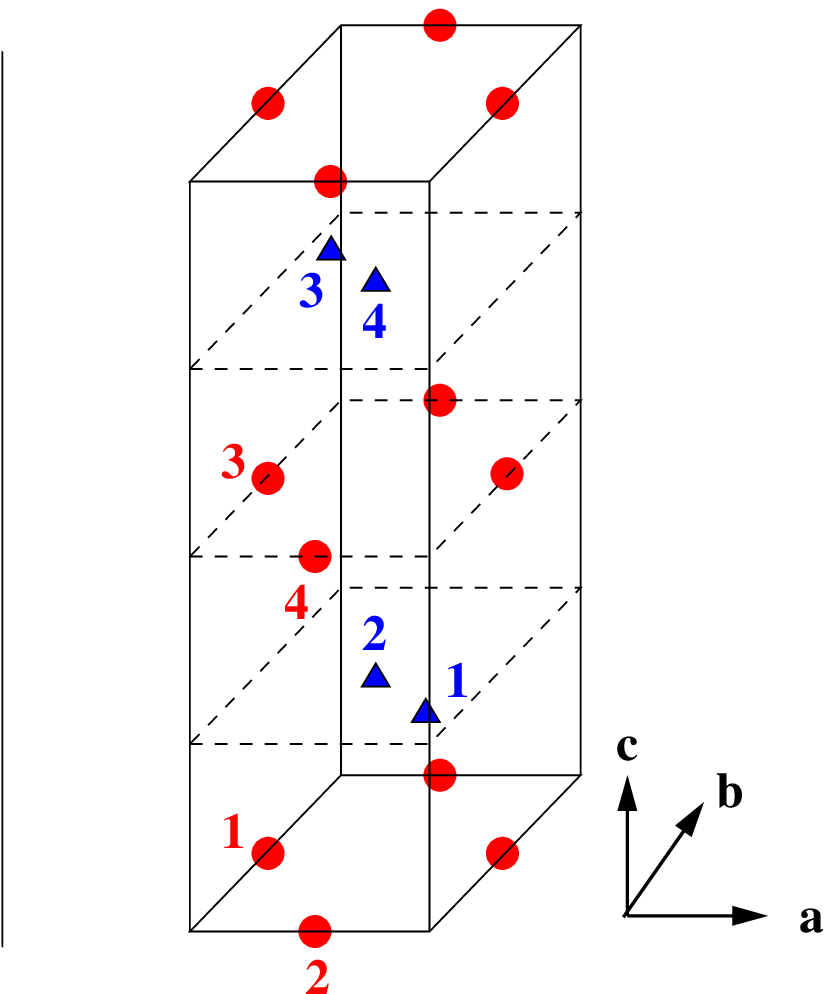}
\caption{Left: The unit cell (ABCDEFGH) showing only Ni ions numbered
as in Table \protect{\ref{SITES}}.  The \protect{\underline b}
axis is perpendicular to the plane of the paper. Dashed circles
represent spins in adjacent planes displaced from the solid symbols
by $\pm {\bf b}/2$.
Right: The unit cell of TbMnO$_3$ showing only the Mn ions (filled circles)
and the Tb ions (filled triangles), numbered as in Table \protect{\ref{SITES}}.}
\label{STR}
\end{figure}
\end{center}

There is one final refinement of our toy model which we must consider,
namely the structure of the magnetic unit cell. In the toy model
considered above, the entire spin structure is characterized by a wavevector
and a single complex vector amplitude. However, more generally, the wavevector
determines only how the spin wavefunction evolves from one unit cell to
the next.  Now we consider how the structure of the wavefunction
within a unit cell is restricted by the symmetry of the crystal lattice.
As a preliminary, we start by discussing the crystal structure of the
two systems, NVO and TMO.  In Table \ref{XTAL}
we list the general equivalent positions which define the space
group operations. For NVO we choose the generators of the space
group to be the identity, $E$, a two-fold rotation about the
$x$ axis, $(x,y,z) \rightarrow (x,\overline y , \overline z)$,
the $x-y$ glide plane, $(x,y,z) -> (x, y+\oh, \overline z + \oh)$,
spatial inversion, $(x,y,z) \rightarrow 
(\overline x, \overline y, \overline z)$, and translations.
For TMO the generators are taken to be $E$, a mirror $z$-plane,
$(x,y,z) \rightarrow (x,y,\oh -z)$, the $y-z$ glide plane,
$(x,y,z) \rightarrow (\oh -x, y+\oh ,z)$, spatial inversion,
and translations.  The magnetic sites are at the positions listed
in Table \ref{SITES} and shown in Fig. \ref{STR}.

\begin{table}
\caption{General positions of the space groups for NVO (top), Cmca
(\#64 in Ref. \protect{\onlinecite{HAHN}}) and for TMO (bottom), Pbnm
(\#62 in Ref. \protect{\onlinecite{HAHN}}). For Cmca the primative
translation vectors are ${\bf a}_1 = (a/2)\hat i + (b/2) \hat j$;
${\bf a}_2= (a/2) \hat i - (b/2) \hat j$; and ${\bf a}_3=c\hat k$.
For Pbnm they are ${\bf a}_1=a\hat i$; ${\bf a}_2 = b\hat j$, and
${\bf a}_3 = c\hat k$.}

\vspace{0.2 in}
\begin{tabular} {|| c | c | c | c||}
\hline \hline
${\bf r}=(x,y,z)$  & $2_z {\bf r}=(\overline x, \overline y+\oh ,z+\oh )$ &
$2_y {\bf r}=(\overline x,y+\oh ,\overline z+\oh )$ & 
$2_x{\bf r}=(x,\overline y,\overline z)$ \\
${\cal I} {\bf r}=(\overline x,\overline y,\overline z)$ &
$m_{xy}{\bf r}=(x,y+\oh ,\overline z+\oh )$ &
$m_{xz}{\bf r}=(x,\overline y+\oh ,z+\oh )$ &
$m_{yz}{\bf r}=(\overline x,y,z)$  \\
\hline \hline
\end{tabular}

\vspace{0.2 in}
\begin{tabular}{||c|c|c|c||}
\hline \hline
${\bf r}=(x,y,z)$ & $m_{xy}{\bf r}=(x,y,\oh-z)$ &
$2_x{\bf r}=(x+ \oh, \overline y+ \oh , \overline z )$ &
$m_{xz}{\bf r}=(x+ \oh, \overline y + \oh, \oh + z)$ \\
${\cal I} {\bf r}=(\overline x , \overline y , \overline z )$ &
$2_z{\bf r}=(\overline x, \overline y,\oh+z)$ &
$m_{yz}{\bf r}=(\oh -x, y +\oh,z)$ &
$2_y{\bf r}=(\oh -x, y + \oh, \oh - z)$ \\
\hline \hline
\end{tabular}
\label{XTAL}
\end{table}

\begin{table}
\caption{Left: Unit cell lattice positions in NVO of
the ${\rm Ni^{2+}}$ ions carrying $S$=$1$
(given as fractions of the cell dimensions $a$, $b$, and $c$).
The ``spine'' sites are $r_n^s$ and the cross-tie sites are
$r_n^c$.  Right: Positions of the Mn and Tb ion sites in the unit cell of
TMO as fractions of the cell sides $a$, $b$, and $c$.}

\vspace{0.2 in}
\begin{tabular}{||c|c||}
\hline \hline
$r^s_1$  &  (0.25, $-0.13$,0.25)\\
$r^s_2$  &  (0.25, 0.13, 0.75)\\
$r^s_3$  &  (0.75, 0.13, 0.75)\\
$r^s_4$  &  (0.75, $-0.13$, 0.25)\\
\hline
$r^c_1$  &  (0,    0,    0)\\
$r^c_2$  &  (0.5,  0,    0.5)\\
\hline \hline
\end{tabular}
\vspace{0.2 in}
\begin{tabular} {|| c || c | c | c|c||}
\hline \hline
& $n=1$ & $n=2$& $n=3$ & $n=4$ \\
\hline
\ Mn \ &\ \ (0,$\oh$ ,0) & $\ \ (\oh  , 0,0 )\ \ $
& $\ \ (0, \oh , \oh )\ \ $ & $\ \ (\oh , 0 , \oh )\ \ $ \\
\hline
\ Tb$^{\rm a}$ \ &\ \ $(x, y, \of )$\ \ &\ \ 
$(x+\oh  , \overline y + \oh , \tf )$\ \
&\ \ $(\overline x , \overline y , \tf )$\ \ &
\ \ $(\overline x + \oh , y+\oh , \of )$\ \ \\
\hline \hline
\end{tabular}
\label{SITES}

\vspace{0.2 in} \noindent
a) $x=0.9836$ and $y=0.0810$ from Ref. \protect{\onlinecite{TMOSTR}}.
\end{table}

\subsection{Representation Theory}

If there are $n_u$ spins in a unit cell, then an incommensurate
state will be described by a wavevector $\bf q$ and the
complex-valued Fourier amplitudes ${\cal S}_{\alpha, \tau} ({\bf q})$,
where $\alpha=x,y,z$ and $\tau=1, 2, \dots n_u$ , in terms of which 
we write the spin wavefunctions in the form
\begin{eqnarray}
S_{\alpha , \tau} ({\bf R}) &=& {\cal S}_{\alpha \tau}({\bf q})
e^{i {\bf q} \cdot ({\bf R} + {\bf r}_{\tau})} 
+ {\cal S}_{\alpha \tau}({\bf q})^*
e^{-i {\bf q} \cdot ({\bf R} + {\bf r}_{\tau})} \ .
\label{PHASEEQ} \end{eqnarray}
where ${\bf r}_{\tau}$ is the position of the $\tau$th site within
the unit cell.  For NVO
$\tau = 1,s$; $2,s$; $3,s$; or $4,s$ for spine (s) sites and $1,c$, or $2,c$
for cross-tie (c) sites.  Note that the complex
amplitudes ${\cal S} (q)$ are defined relative to
the phase, ${\bf q} \cdot ({\bf R}+{\bf r}_\tau )$ which would
obtain if the wave were perfectly sinusoidal.  (This convention
will simplify later results.)

We now discuss how symmetry restricts the possible values of the
amplitudes ${\cal S}_{\alpha, \tau}(q)$ and how these variables are
determined via diffraction experiments.  The analysis of the symmetry
of such systems in terms of their point groups is not developed.
Accordingly, a model-independent (representation) analysis\cite{WILLS}
is customarily invoked in such cases.  {\it If the magnetic ordering transition
is assumed to be continuous}, then the phase transition is signalled by
an instability in the quadratic terms when the free energy is
expanded in powers of the order parameters.
In that case, the spin ordering within a unit cell will be determined
by the critical eigenvector associated with the first eigenvalue
of the quadratic free energy which passes through zero as the temperature
is lowered. This phenomenon has been discussed previously in connection with
wavevector selection. As we shall see, this analysis of the magnetic
symmetry is essential to construct the allowed couplings
between magnetic and ferroelectric order parameters.

In order to conserve wavevector the quadratic terms in the Landau expansion
associated with the selected wavevector $\bf q$ must be of the form
\begin{eqnarray}
f_2 &=& \sum_{\alpha , \beta , \tau , \tau'}
c_{\alpha , \tau ; \beta , \tau'}({\bf q}) {\cal S}_{\alpha , \tau} (-{\bf q})
{\cal S}_{\beta , \tau'} ({\bf q})  \ ,
\label{F2EQ} \end{eqnarray}
where ${\cal S}_{\alpha , \tau} (-{\bf q}) = {\cal S}_{\alpha , \tau} ({\bf q})^*$.
For $f_2$ to be real for any choice of the complex-valued Fourier amplitudes,
it is required that
$c_{\alpha , \tau; \beta , \tau'}({\bf q})^*=c_{\beta , \tau';\alpha ,\tau }({\bf q})$.
As in the case of phonons or other normal modes,
the eigenvectors of this quadratic form
can be labeled according to the irreducible representations (irreps)
$\Gamma_n$ of the paramagnetic phase which leave the wavevector invariant.
(This group of symmetry operations is called the group of the
wavevector or the ``little group.'' The relevant symmetry is that of the
paramagnetic phase because the expansion of the free energy in
powers of the order parameters is relative to this phase.\cite{PSYM1,PSYM2})
For the orthorhombic systems NVO and TMO considered
in this review, all the irreps are one dimensional.  So essentially,
the eigenvectors must be even or odd under the rotation or
mirror (or glide) operations of the little group. 

The result of the group theoretical analysis is that one expresses
the Fourier amplitudes ${\cal S}_{\alpha \tau}({\bf q})$ in terms
of {\it symmetry adapted coordinates} $m^{(n)}_\eta$,
associated with the irrep $\Gamma_n$.  Here the subscript $\eta$ indicates
the spin component and type of site within the unit cell.  For instance,
for NVO, (as we will see later in Table \ref{NVOSPIN}),
for $n=3$ or $n=4$, $\eta$ ranges over five values, three
associated with spin components on spine sites and two associated with
spin components on cross tie sites.  For $n=1$ or $n=2$, $\eta$ ranges over
four values, three associated with spin components on spine sites and one
associated with a cross tie spin component.
If we write
\begin{eqnarray}
{\cal S}_{\alpha \tau}(q) &=& \sum_{n, \eta} U^{(n)}_{\alpha \tau; \eta}
m^{(n)}_\eta (q) \ ,
\label{UEQ} \end{eqnarray}
then $U^{(n)}_{\alpha \tau; \eta}$ is the $\eta$th {\it symmetry adapted
basis function} of irrep
$n$ in the sense that it specifies the $\alpha$ spin component for the
site $\tau$ in the unit cell. These basis functions are given
in Tables \ref{NVOSPIN}, \ref{MNSPIN}, and \ref{TBSPIN}, below.
(Thus, for condensation of spins in NVO via irrep $\Gamma_4$, Table IV
indicates that the specification of the magnetic structure requires
fixing the five complex-valued symmetry adapted coordinates $m_{sx}^{(4)}$,
$m_{sy}^{(4)}$, $m_{sz}^{(4)}$, $m_{cy}^{(4)}$, and $m_{cz}^{(4)}$.)
The advantage of this formalism is that the quadratic free energy only
couples symmetry adapted coordinates having the same irrep superscript $n$:
\begin{eqnarray}
f_2 &=& \sum_n \sum_{\eta \eta'} F_{\eta , \eta'}^{(n)}
m_\eta^{(n)}(q)^* m_{\eta'}^{(n)}(q) \ ,
\label{F22EQ} \end{eqnarray}
where the reality of $f_2$ requires that ${F^{(n)}_{\eta, \eta'}}^*
= F^{(n)}_{\eta' ,\eta}$.

To deal with this free energy it is useful to introduce {\it normal
coordinates} $\xi^{(n)}_\rho$ in terms of which the quadratic form
is diagonal.  So we set
\begin{eqnarray}
m^{(n)}_\eta (q) = \sum_\rho u^{(n)}_{\eta \rho} \xi^{(n)}_\rho \ ,
\label{uEQ} \end{eqnarray}
and the unitary matrix ${\bf u}^{(n)}$ (which diagonalizes the
quadratic free energy) is chosen so that
\begin{eqnarray}
f_2 &=& \sum_{n, \rho} \lambda_\rho^{(n)} |\xi^{(n)}_\rho|^2 \ .
\end{eqnarray}
Usually 
\begin{eqnarray}
\lambda_\rho^{(n)}=dkT - \tilde \lambda_\rho{(n)}\ ,
\label{LAMOFT} \end{eqnarray}
where $\tilde \lambda_\rho^{(n)}$ is a temperature-independent interaction
term.  As we shall see, the transformation ${\bf U}$ to symmetry adapted
coordinates is determined by the symmetries of the system, whereas the
further transformation to normal coordinates depends on the details
of the interactions between spins.  However, the
explicit form of ${\bf u}$ will not be needed for our analysis. 

To make this analysis more intuitive we may liken it to
the problem of a particle moving in a spherically symmetric potential.
For illustrative purposes we consider a Coulomb potential with a weak
spherically symmetric perturbation.  To solve this problem,
one introduces symmetry adapted basis functions for the various
irreps, which in this case are {\it s} functions $\psi^{(s)}_n$,
{\it p} functions $\psi^{(p_x)}_n$, $\psi^{(p_y)}_n$, $\psi^{(p_z)}_n$,
{\it etc.} Then $\Psi_n^{(q)}$, the $n$th eigenfunction of type $q$
($q=s$, $q=p_x$, {\it etc}.) is written in terms of the basis functions as
\begin{eqnarray}
\Psi^{(q)}_n &=& \sum_m u^{(q)}_{n \rho} \psi^{(q)}_\rho \ .
\end{eqnarray}
The basis functions $\psi_m^{(q)}$ are constructed solely from symmetry
arguments.  The actual value of the transformation ${\bf u}$ to
eigenfunctions depends on the details of the potential. 

We now return to the problem of incommensurate magnets.
At high temperature, i. e. in
the paramagnetic phase all the eigenvalues $\lambda^{(n)}_\rho$
are positive and the trial free energy is minimized by setting
$\xi_\rho^{(n)}=0$ for all $\rho$ and $n$, so that
all the $S_{\alpha \tau}({\bf R})$ are zero: the
paramagnetic phase is stable against the formation
of long range magnetic order.  As the temperature is
lowered, one of these eigenvalues will pass through zero [c. f.
Fig. \ref{CHIFIG} and Eq. (\ref{LAMOFT})] and
the irrep for which this happens becomes ``active," so to speak.
It is conceivable that eigenvalues $\lambda^{(n)}_\rho$
and $\lambda^{(n')}_{\rho'}$ of two different irreps could be
degenerate.  We reject the possibility of such an accidental
degeneracy.  However, if one adjusts an additional control parameter,
such as the pressure, it is possible to reach a higher order
critical point where two irreps simultaneously become active.
A simple example of this principle arises when one treats a
ferromagnet on a tetragonal lattice.  In that case one irrep
is one dimensional and corresponds to the ferromagnetic order
lying along the four-fold crystal ($\underline c$) axis and
the other irrep is two-dimensional and corresponds to ordering in
the plane perpendicular to the $\underline c$ axis.  Clearly, the
mean-field transition temperatures for these two distinct
orderings should be assumed to be different.  If the anisotropy
is easy-axis the ferromagnetic moment will lie along the
$\underline c$ axis and if the anisotropy is easy plane the
moment will be perpendicular to the $\underline c$ axis.
It is possible for the anisotropy to vanish, but only by
adjusting another thermodynamic variable, such as uniaxial stress.
One therefore concludes that criticality is associated with a
{\it single} irrep.  Since the transformation to symmetry adapted
coordinates can be determined using only symmetry considerations,
the possible patterns of spin ordering within the unit cell are
strongly restricted. 

In the usual presentation of representational
analysis,\cite{WILLS} the only constraint on the symmetry adapted
coordinates is that they transform properly under the operations of the
little group, $g_k$.  For NVO the wavevector in
question is along the crystal \underline a axis, and the generators
of $g_k$ are $2_x$, a two-fold rotation about the $x$-axis
(we often refer to \underline a, \underline b, and \underline c as
$x$, $y$, and $z$, respectively) and
$m_{xy}$, a glide operation which takes $z$ into $-z$
followed by a translation of $({\bf b+c})/2$.
For TMO the wavevector is along the crystal \underline b axis and the
generators of $g_k$ are a two-fold screw rotation about the $y$-axis,
$2_y$ and the mirror plane $m_{xy}$.
These operations are defined in Table \ref{XTAL}.
It should be noted that in the case of TMO the four symmetry
operations do not actually form a ``group" because $[2_y]^2$
picks up a phase factor and is thus {\it not} equal to a 
member of the group. This situation also occurs in connection
with the application of group theory to the band structure
of nonsymmorphic space groups.\cite{GROUP,CORNWELL}
(These are space groups for which some pure point group
operations are not space group operations.)
The formal solution to this problem is cumbersome.\cite{ZAK}
Here we may avoid these complications by defining the operator
$\tilde 2_y \equiv e^{iqa/2} 2_y$, so that 
$[\tilde 2_y]^2$ translates the wave into itself.
The character tables for these little groups are given in Table \ref{CHAR}.
(In essence, the character table tells whether wavefunctions associated
by $\Gamma_n$ are even or odd under the symmetry operations listed.)

\begin{table}
\caption{Character Table for the symmetry operations ${\cal O}$
of the group$^{(\rm a)}$ $G_{\bf k}$ for the irreps $\Gamma_n$ for
incommensurate magnetic structure of (left) NVO with ${\bf k}=(q,0,0)$
and (right) TMO for ${\bf k}=(0,q,0)$.}

\vspace{0.2 in}
\begin{tabular}{||c|ccccc||}
\hline \hline
${\cal O}^{({\rm b})}=$ &$1$ &  $2_x$ & $m_{xy}$ & $m_{xz}$ & \\ \hline
$\Gamma^1$ & 1 &  1 &  $1$ &  $1$ &\\
$\Gamma^2$ & 1 &  1 & $-1$ & $-1$ &\\
$\Gamma^3$ & 1 & $-1$ &  $-1$ & $1$ &\\
$\Gamma^4$ & 1 & $-1$ & $1$ &  $-1$ &\\ 
\hline \hline
\end{tabular}
\hspace{1 in} \begin{tabular} {|| c || c | c | c | c ||} 
\hline \hline
${\cal O}^{({\rm a})}=$ \ & $E$ & $\tilde 2_y$ & $m_{xy}$
& $\tilde m_{yz}$ \\
\hline \hline
$\Gamma_1$\ \ & 1 \ \ & 1 \ \ & $1$ \ \ & $1$ \ \\
$\Gamma_2$\ \ & 1 \ \ & 1 \ \ & $-1$ \ \ & $-1$ \ \\
$\Gamma_3$\ \ & 1 \ \ & $-1$ \ \ & $1$ \ \ & $-1$ \ \\
$\Gamma_4$\ \ & 1 \ \ & $-1$ \ \ & $-1$ \ \ & $1$ \ \\
\hline \hline
\end{tabular}

\vspace{0.2 in}
\noindent
a)  For an operator ${\cal O}$ we define $\tilde {\cal O}=
e^{iqa/2} {\cal O}$.

\noindent
b) Operators (without tildes) are defined in Table \ref{XTAL}.
\label{CHAR}
\end{table}
\vspace{0.2 in}

For either NVO or TMO the next step is to construct the symmetry
adapted basis functions which transform according to the irreps
listed in Table \ref{CHAR}.
These allowed symmetry-adapted spin functions are listed in Table
\ref{NVOSPIN} for NVO and in Tables \ref{MNSPIN} and \ref{TBSPIN}
for TMO.  (In Table \ref{MNSPIN} the symmetry adapted coordinates
are denoted $M^{\rm M}_\alpha$ and in Table \ref{TBSPIN} they are denoted
$M^{\rm T,m}_\alpha$.)  Note that each symmetry adapted coordinate
appearing in these
tables is complex-valued, each with an amplitude and (within the
analysis discussed up to now) an independent phase.  Note also that
simply specifying the symmetry does not fix all the spin components
in the unit cell.  Rather it allows a choice of spin components for
each set of symmetry-equivalent sites. For NVO one can
specify the spin components of a single spine site and those of a 
single cross-tie site (unless a component is forced by symmetry to be
zero).  Having done this, the spin components
of the other spine and and cross-tie sites are then fixed by the
symmetry properties of the irrep in question.

When the temperature is lowered further and the LTI phase is
entered, then an additional irrep will become active
via a second continuous phase transition. For NVO the new LTI
representation (in addition to $\Gamma_4$ already present in the
HTI phase) is\cite{PRL} $\Gamma_1$ and for TMO the new LTI
representation (in addition to $\Gamma_3$ already present in the
HTI phase) is\cite{TMO} $\Gamma_2$. In an appendix we discuss
that when two different irreps are active, their presence does
not induce the development of a third irrep.  However, had there
been a further phase transition from the LTI phase into yet another
incommensurate phase with three irreps, then the presence of three
different irreps would induce the presence of a fourth one.

\vspace{0.3 in}
\begin{table}[ht]
\caption{Symmetry adapted basis functions $U^{(n)}_{\alpha \tau; \eta}$
which transform according to the irreducible representation $\Gamma_n$
for the incommensurate phase associated
with ${\bf k}=(q,0,0)$ for the ${\rm Ni}$ spine (s) and cross-tie (c)
sites of NVO. Here ${\cal S}_{1s}$ denotes the vector with components
${\cal S}_{x,1s}$, ${\cal S}_{y,1s}$, ${\cal S}_{z,1s}$, and similarly
for the other ${\cal S}$'s.  The symmetry adapted  coordinates
$m_{r \alpha}^{(n)}$ assume complex values, as discussed in
the text. The numbering of sites is given in Table~\protect\ref{SITES}.
The phase factors of $i$ are chosen to simplify the transformation
properties under spatial inversion, as is discussed below.}

\vspace{0.2 in}
\begin{tabular}{||c||c|c|c|c||}
\hline \hline
&${\bf U}^{(1)}$ & ${\bf U}^{(2)}$ & ${\bf U}^{(3)}$ & ${\bf U}^{(4)}$
\vspace{0.05cm}\\
\hline
${\cal S}_{1s}$ & $(im_{sx}^{(1)}, m_{sy}^{(1)}, im_{sz}^{(1)})$ 
& $(m_{sx}^{(2)}, im_{sy}^{(2)},m_{sz}^{(2)})$
& $(im_{sx}^{(3)} ,m_{sy}^{(3)}, im_{sz}^{(3)})$
& $(m_{sx}^{(4)} , im_{sy}^{(4)} ,m_{sz}^{(4)})$ \\ \hline
${\cal S}_{2s}$ & $(im_{sx}^{(1)} ,-m_{sy}^{(1)} ,-im_{sz}^{(1)})$
& $(m_{sx}^{(2)} , -im_{sy}^{(2)} ,-m_{sz}^{(2)}) $
& $(-im_{sx}^{(3)} ,m_{sy}^{(3)} ,im_{sz}^{(3)})$
& $(-m_{sx}^{(4)} , im_{sy}^{(4)} ,m_{sz}^{(4)})$ \\ \hline
${\cal S}_{3s}$ & $(-im_{sx}^{(1)} , \, m_{sy}^{(1)} , -im_{sz}^{(1)})$
& $( \, m_{sx}^{(2)} , - im_{sy}^{(2)} , \, m_{sz}^{(2)} ) $
& $(- im_{sx}^{(3)} , \, m_{sy}^{(3)} , - im_{sz}^{(3)})$
& $( \, m_{sx}^{(4)} - im_{sy}^{(4)} , \, m_{sz}^{(4)})$ \\ \hline
${\cal S}_{4s}$ & $ (- im_{sx}^{(1)} , - m_{sy}^{(1)} , im_{sz}^{(1)})$
& $(\, m_{sx}^{(2)} , \, im_{sy}^{(2)} , - m_{sz}^{(2)})$
& $(\, im_{sx}^{(3)} ,  m_{sy}^{(3)} , - im_{sz}^{(3)})$
& $(- m_{sx}^{(4)} , - im_{sy}^{(4)} , \, m_{sz}^{(4)} )$ \\ \hline
${\cal S}_{1c}$ & $(m_{cx}^{(1)} ,0 ,0)$ & $(m_{cx}^{(2)} ,0 ,0)$&
$(0 ,m_{cy}^{(3)} ,m_{cz}^{(3)})$ & $(0 ,m_{cy}^{(4)} ,m_{cz}^{(4)})$
\\ \hline ${\cal S}_{2c} $ & $(- m_{cx}^{(1)} , 0 ,0)$
& $(\, m_{cx}^{(2)} , 0 ,0)$& $(0 , m_{cy}^{(3)} , -m_{cz}^{(3)})$
& $(0 ,-m_{cy}^{(4)} , m_{cz}^{(4)})$ \\
\hline \hline
\end{tabular}
\label{NVOSPIN}
\end{table}

\begin{center}
\begin{figure}
\includegraphics[width=12cm]{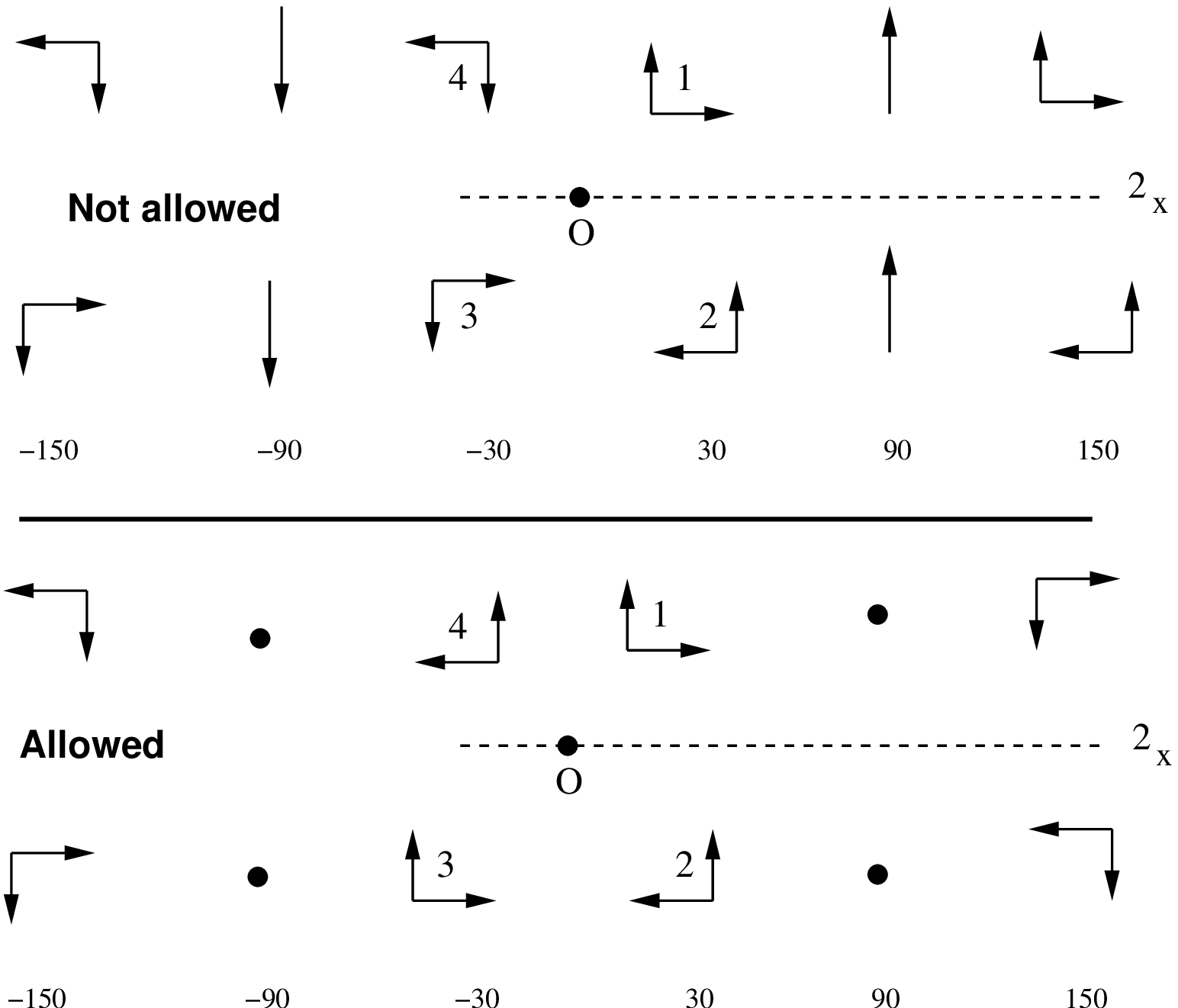}
\caption{Two structures for spine spins in an \underline a -\underline c plane
for NVO obtained using spin components from Table
\protect{\ref{NVOSPIN}} which transform according to irrep $\Gamma_4$
for wavevector $k=2 \pi /(3a)$, where $a/2$ is the distance between
sites.  The axis for the two-fold rotation $2_x$ is indicated.  The
glide plane, which relates spins in adjacent layers is not shown.
The sites are numbered as in Table \ref{SITES}.  At the
bottom of each panel we give the phase (in degrees) of the wave at each site.
Top: for $m_{sx}^{(4)}=1$, $m_{sz}^{(4)}=-i$ and the other
parameters equal to zero.  Bottom: for $m_{sx}^{(4)}=m_{sz}^{(4)}=1$
and the other parameters equal to zero. As we shall see in Eq. 
(\protect{\ref{CRITICAL}}), the order parameters $m_{sx}^{(4)}$ and 
$m_{sz}^{(4)}$ must have the same phase, {\it i. e.}
$m_{sx}^{(4)}/m_{sz}^{(4)}$ must be real.  Only the bottom configuration
satisfies this constraint.}
\label{ILLUS}
\end{figure}
\end{center}

As an illustration of how to apply the above results, we show, in Fig.
\ref{ILLUS}, typical spin  configurations for NVO which result from
the spin wavefunctions which transform according to irrep \#4.
The configuration shown in the top panel is not allowed because,
as we discuss below, it does not respect inversion symmetry.

\begin{table}
\caption{Symmetry adapted basis functions $U^{(n)}_{\alpha \tau; \eta}$
which transform according to the irreducible representation $\Gamma_n$
for the incommensurate phase associated with ${\bf k}=(q,0,0)$ for the
Mn sites in TMO.  The complex-valued symmetry adapted coordinates are
here denoted $M_\alpha^{\rm M}$. The sites are numbered as in Table
\protect{\ref{SITES}}.}

\vspace{0.2 in}
\begin{tabular} {|| c || c | c | c | c ||} 
\hline \hline
Mn Site & ${\bf U}^{(1)}$ & ${\bf U}^{(2)}$ & ${\bf U}^{(3)}$
& ${\bf U}^{(4)}$ \\
\hline \hline
\ \ $1$\ \  &\ \ $(M^{\rm M}_x , M^{\rm M}_y, M^{\rm M}_z)$\ \ &
\ \ $(M^{\rm M}_x , M^{\rm M}_y , M^{\rm M}_z)$
\ \ &\ \ $(M^{\rm M}_x, M^{\rm M}_y M^{\rm M}_z)$\ \ &
\ \ $(M^{\rm M}_x, M^{\rm M}_y, M^{\rm M}_z)$\ \ \\
$2$ & $( M^{\rm M}_x , -M^{\rm M}_y, -M^{\rm M}_z)$ &
$(-M^{\rm M}_x, M^{\rm M}_y, M^{\rm M}_z)$ &
$(-M^{\rm M}_x, M^{\rm M}_y, M^{\rm M}_z)$ &
$(M^{\rm M}_x, -M^{\rm M}_y, -M^{\rm M}_z)$ \\
$3$ & $(-M^{\rm M}_x, -M^{\rm M}_y, M^{\rm M}_z)$ &
$(M^{\rm M}_x, M^{\rm M}_y, -M^{\rm M}_z)$ &
$(-M^{\rm M}_x, -M^{\rm M}_y, M^{\rm M}_z)$ &
$(M^{\rm M}_x, M^{\rm M}_y, -M^{\rm M}_z)$ \\
$4$ & $(-M^{\rm M}_x, M^{\rm M}_y, -M^{\rm M}_z)$ &
$(-M^{\rm M}_x, M^{\rm M}_y, -M^{\rm M}_z)$ &
$(M^{\rm M}_x, -M^{\rm M}_y, M^{\rm M}_z)$ &
$(M^{\rm M}_x, -M^{\rm M}_y, M^{\rm M}_z)$ \\
\hline \hline
\end{tabular}
\noindent
\label{MNSPIN}
\end{table}

\begin{table}
\caption{Symmetry adapted basis functions $U^{(n)}_{\alpha \tau; \eta}$
which transform according to the irreducible representation $\Gamma_n$
for the incommensurate phase associated with ${\bf k}=(q,0,0)$ for the
Tb sites in TMO.  The complex-valued symmetry adapted coordinates are
here denoted $M_\alpha^{{\rm T},n}$. The sites are numbered as in Table
\protect{\ref{SITES}}.}

\vspace{0.2 in}
\begin{tabular} {|| c || c | c | c | c ||} 
\hline \hline
Tb Site & ${\bf U}^{(1)}$ & ${\bf U}^{(2)}$ & ${\bf U}^{(3)}$
& ${\bf U}^{(4)}$ \\
\hline \hline
\ \ $1$\ \  &
\ \ $(0, 0, M^{\rm T,2}_z)$\ \ &
\ \ $(M^{\rm T,2}_x,M^{\rm T,2}_y,0)$ \ \ &\ \ $(0,0,M^{\rm T,2}_z)$\ \ &
\ \ $(M^{\rm T,2}_x, M^{\rm T,2}_y, 0)$\ \ \\
$2$ &
\ \ $(0, 0, -M^{\rm T,1}_z)$\ \ &
\ \ $(-M^{\rm T,1}_x,M^{\rm T,1}_y,0)$ \ \ &\ \ $(0,0,M^{\rm T,1}_z)$\ \ &
\ \ $(M^{\rm T,1}_x, -M^{\rm T,1}_y, 0)$\ \ \\
$3$ &
\ \ $(0, 0, M^{\rm T,1}_z)$\ \ &
\ \ $(M^{\rm T,1}_x,M^{\rm T,1}_y,0)$ \ \ &\ \ $(0,0,M^{\rm T,1}_z)$\ \ &
\ \ $(M^{\rm T,1}_x, M^{\rm T,1}_y, 0)$\ \ \\
$4$ &
\ \ $(0, 0, -M^{\rm T,2}_z)$\ \ &
\ \ $(-M^{\rm T,2}_x,M^{\rm T,2}_y,0)$ \ \ &\ \ $(0,0,M^{\rm T,2}_z)$\ \ &
\ \ $(M^{\rm T,2}_x, -M^{\rm T,2}_y, 0)$\ \ \\
\hline \hline
\end{tabular}
\vspace{0.3 in}
\label{TBSPIN}
\end{table}

\subsection{Effect of Inversion Symmetry for NVO}

Up to now we only used the consequences of the symmetry of the group
of transformations which leave the wavevector invariant.
However, as we have observed previously, the quadratic free energy of Eq.
(\ref{F2EQ}) or of Eq. (\ref{F22EQ})
must be invariant under all the operations of the paramagnetic
space group.  In particular, the operations for the systems we consider here
which are {\it not} in the little group of the wavevector are those generated
by spatial inversion ${\cal I}$.
We now discuss how inversion symmetry places restrictions on
symmetry adapted coordinates.  Usually when one introduces
an additional symmetry, the matrix for the quadratic free energy
becomes block diagonal.  Here, the result of the additional
symmetry is not to reduce the size of the submatrices for the
quadratic free energy, but rather it places additional constraints
on the transformation matrix ${\bf u}$ of Eq. (\ref{uEQ}).  As we shall
see, when ${\bf U}$ is appropriately defined, then inversion symmetry
restricts all the matrix elements $F^{(n)}_{\eta, \eta'}$ to be real.
Consequently ${\bf u}$ is real, apart from an overall phase
$\phi_n$ which is associated with each irrep $\Gamma_n$.

We first analyze the situation for NVO.
We need to determine the effect of ${\cal I}$ on the spin wavefunctions
listed in Table \ref{NVOSPIN}. Recall that the magnetic moment is a
pseudovector.  That means that under spatial inversion the orientation
of the moment is unchanged, but it is simply transported
from its initial location at ${\bf r}$ to the transformed location, $-{\bf r}$.
Looking at Eq. (\ref{PHASEEQ}) we see that spatial inversion interchanges
${\cal S}$ and ${\cal S}^*$, but since the orientation is unchanged, spatial
inversion will not affect the spin component label $\alpha$.  However,
spatial inversion does interchange Ni sublattices \#1 and \#3 and also \#2
and \#4.  In other words, in the notation of Table \ref{NVOSPIN}, we have
\begin{eqnarray}
{\cal I} {\cal S}_{x,s1} = {\cal S}_{x,s3}^* \ , \ \ \ 
{\cal I} {\cal S}_{x,s2} = {\cal S}_{x,s4}^* \ .
\label{INVERSE} \end{eqnarray}
To see the consequences of these relations, consider the effect of
the first of these two relations acting on the basis functions of
${\bf U}^{(1)}$, for instance.  This relation is
\begin{eqnarray}
{\cal I} (im_{sx}^{(1)}, m_{sy}^{(1)},im_{sz}^{(1)}) =
(-im_{sx}^{(1)}, m_{sy}^{(1)},-im_{sz}^{(1)})^* \ ,
\end{eqnarray}
which can be written as
\begin{eqnarray}
{\cal I} (m_{sx}^{(1)}, m_{sy}^{(1)},m_{sz}^{(1)}) =
(m_{sx}^{(1)}, m_{sy}^{(1)},m_{sz}^{(1)})^* \ .
\end{eqnarray}
This same analysis can be repeated for the other representations and
also for the second relation of Eq. (\ref{INVERSE}). Then we see that
the choices of the phase factors $i$ in Table \ref{NVOSPIN} leads to
the simple result that for
$\alpha=x,y,$ or $z$, and independent of representation $\Gamma_n$
\begin{eqnarray}
{\cal I} {\bf m}_{s\alpha}^{(n)} (q)  &=& {\bf m}_{s\alpha}^{(n)} (q)^* \ .
\label{NVOIEQ} \end{eqnarray}
For the cross tie sites, the situation is similar except that under spatial
inversion each sublattice is transformed into itself.  Thus, we 
find that
\begin{eqnarray}
{\cal I} {\bf m}_{c\alpha}^{(n)} (q) = {\bf m}_{c\alpha}^{(n)} (q)^* \ .
\label{NVOIIEQ} \end{eqnarray}
So, generally for NVO we have for {\it any representation} $\Gamma_n$
\begin{eqnarray}
{\cal I} m_\eta^{(n)} (q) = m_\eta^{(n)}(q)^* \ ,
\label{NVOINV} \end{eqnarray}
where $\eta =s,\alpha$ or $\eta=c,\alpha$, where $\alpha=x,y,z$.
(Note that this relation does not imply inversion symmetry.
If the system has inversion symmetry about the origin,
then ${\cal I} m_\eta^{(n)} (q) = m_\eta^{(n)}$, and
magnetic order can not induce ferroelectric order.  Thus
one can not have ferroelectric order if all the $m$'s are real.)

Now we return to the Landau expansion of Eq. (\ref{F22EQ}) to see how these
relations affect the determination of the critical eigenvectors. First
consider the situation for NVO when one has Eq. (\ref{NVOINV}).
Recall that
\begin{eqnarray}
F^{(n)}_{\eta , \eta'} = [F^{(n)}_{\eta' , \eta }]^* \ .
\label{HERMIT} \end{eqnarray}
Now we use Eq. (\ref{NVOINV}) to see the consequences of the invariance
of $f_2$ with respect to inversion symmetry.  we again use the
fact that $f_2$ must be invariant under the symmetry operations of
the paramagnetic phase,\cite{PSYM1,PSYM2} and spatial inversion
is one such symmetry.  We find that
\begin{eqnarray}
{\cal I} f_2 &=& \sum_n \sum_{\eta \eta'}
F^{(n)}_{\eta,\eta'} [{\cal I} m_\eta^{(n)}(q)^*]
[{\cal I} m_{\eta'}^{(n)} (q)] =
\sum_n \sum_{\eta \eta'}
F^{(n)}_{\eta,\eta'} m_\eta^{(n)}(q)
m_{\eta'}^{(n)} (q)^* \nonumber \\ &=&
\sum_n \sum_{\eta \eta'}
F^{(n)}_{\eta',\eta} m_\eta^{(n)}(q)^*
m_{\eta'}^{(n)} (q) \ .
\end{eqnarray}
Thus we see that inversion invariance  of $f_2^{(n)}$ implies that
$F^{(n)}_{\eta , \eta'} = F^{(n)}_{\eta' , \eta}$.
Combining this with Eq. (\ref{HERMIT})
we see that these coefficients must all be real valued.  This means that
all the components of the eigenvectors of the quadratic free energy, when
written in terms of the variables of Eq. (\ref{F22EQ}), can be 
taken to be real valued.  This does not mean that these variables
{\it must} be real.  Rather, since these variables are allowed to be complex,
one may introduce an overall complex phase factor.
So, the critical eigenvector, which we denote $u^{(4)}_{\eta, \rho}$
with $\rho=0$, has an arbitrary
overall phase, in which case the amplitudes in the HTI phase are given as
\begin{eqnarray}
m_\eta^{(4)} &=& \sigma_{\rm HTI} u^{(4)}_{\eta , 0} =
\sigma_{\rm HTI} e^{i \phi_{\rm HTI}} r^{(4)}_{\eta , 0} \ ,
\label{CRITICAL} \end{eqnarray}
in terms of the real-valued eigenvalue components $r_{\eta ,0}^{(4)}$.
Because we have just found that the matrix ${\bf F}^{(n)}$
is real (and symmetric), the components of the eigenvector $r_{\eta , 0}$
are real valued, but, as mentioned above, since they depend on
the details of the interactions, we do not say anything about their
explicit form.  Also, because we have introduced an overall
scale factor $\sigma_{\rm HTI}$, we may require that $\sum_\eta
[r^{(4)}_{\eta ,0}]^2 =1$.  Equation (\ref{CRITICAL}) shows that we
are dealing with an $x-y$ like order parameter
$\sigmav \equiv \sigma_{\rm HTI}e^{i \phi_{\rm HTI}}$
which has an amplitude and a phase. (As the temperature is varied near
$T_{\rm HTI}$, Landau theory gives the approximate result
$\sigma_{\rm HTI} \sim (T_{\rm HTI}-T)^{1/2}$.)
In the appendix this argument (showing that the $u^{(n)}_{\eta, \rho}$
are real apart from an overall phase factor) is extended
to include fourth order terms in the free energy.  In analyzing experimental
data, it is very helpful to realize that apart from the overall phase,
$\phi_{\rm HTI}$, all the phases of the spin amplitudes are fixed.
When speaking in terms of the spin components, ${\cal S}_{\alpha \tau}(q)$,
the listing of Table \ref{NVOSPIN} indicates that (for irrep \#4, for
instance), ${\cal S}_{x,1s}(q)$, ${\cal S}_{z,1s}(q)$, ${\cal S}_{y,1c}(q)$,
and ${\cal S}_{z,1c}(q)$ will all have the same phase, but (due to the
factor $i$), ${\cal S}_{y,1s}(q)$ will be
out of phase with the other variables.  As it happens, unless
a huge number (several thousand) of reflections are monitored, it
is impossible to use diffraction data to fix the relative phases
with any degree of certainty.  Thus, this theoretical development
is useful to completely determine the spin structure of complicated
systems such as NVO or TMO.

We now check to see whether or not the HTI phase has a center
of inversion symmetry, in which case, a spontaneous polarization
can not be induced in this phase.  We will show that a phase with
a single representation has inversion symmetry.  First of all,
because we assume incommensurability, we can redefine the origin
to be arbitrarily close to a lattice site at $R$, such that $\phi-qR$
is a multiple of $2 \pi$.  We have already noted that
${\cal I} m_\alpha^\tau = {m_\alpha^\tau}^*$.  But if $\phi$ is
redefined to be zero, this implies that
${\cal I} m_\alpha^\tau = m_\alpha^\tau$, which means that the
spin structure has inversion symmetry about the redefined origin.
In Fig. \ref{ILLUS} we show an example of a system obeying
Eq. (\ref{CRITICAL}) which does have inversion symmetry and one
having an arbitrary set of parameters out of Table \ref{NVOSPIN}
which does not satisfy Eq. (\ref{CRITICAL}).  This latter structure
does not display inversion symmetry.  Note that, as exemplified by
the bottom panel of Fig. \ref{ILLUS}, it is possible for a
structure to be noncollinear, but to have a center of inversion
symmetry. So noncollinearity, in and of itself, does not
guarantee having a spontaneous polarization.

The analysis of the LTI phase is similar.  Here again one can use
the transformation properties of the order parameters under inversion
to fix the phases of the spin amplitudes.  Again, at quadratic
order, one has the same result as for the HTI phase: all the LTI order
parameters $m_\alpha^s$ for the LTI irrep have the same phase. The
analysis is extended to quartic order in the appendix.

\subsection{Effect of Inversion Symmetry for TMO}

For TMO each Mn sublattice is transformed into itself, so for the
parameters of Table \ref{MNSPIN} we have
\begin{eqnarray}
{\cal I} M^{\rm M}_\alpha (q) =M^{\rm M}_\alpha (q)^*\ .
\end{eqnarray}
For the Tb spins, inversion transforms sublattice \#1 into \#3
and \#2 into \#4, so that for them one has
\begin{eqnarray}
{\cal I} M^{{\rm T},n}_\alpha (q) = M^{{\rm T},3-n}_\alpha (q)^* \ ,
\end{eqnarray}
where $n=1,2$ and $\alpha = x, y$, or $z$. [${\rm M}_\alpha^{{\rm T},1} (q)$
is associated with sites \#1 and \#4 and ${\rm M}_\alpha^{{\rm T},2} (q)$
is associated with sites \#2 and \#3.]

The situation for TMO is slightly more complicated than it was for NVO
because of the presence of the lower-symmetry Tb sites.
In the HTI phase one can repeat the argument used for NVO to show that all
the spin components on the Mn sites, $M_\alpha^{\rm M}$, have the same phase.
In the HTI phase the irrep for TMO was determined to be $\Gamma_3$.
Accordingly we study the quadratic free energy $f_2^{(3)}$ associated with
a single irrep, $\Gamma_3$.  In matrix notation we have the quadratic free
energy in terms of symmetry adapted coordinates as
\begin{eqnarray}
f_2^{(3)} &=& \left[ M_x^* , M_y^* ,  M_z^*  , T_1^* , T_2^* \right]
\left[ \begin{array} {c c c c c}
a & b & c & z_1 & z_2 \\
b & d & e & z_3 & z_4 \\
c & e & f & z_5 & z_6 \\
z_1^* & z_3^* & z_5^* & g & z_7 \\
z_2^* & z_4^* & z_6^* & z_7^* & h \\
\end{array} \right] \left[ \begin{array} {c}
M_x \\ M_y \\ M_z \\ T_1 \\ T_2 \\ \end{array}  \right] \ .
\label{MATRIX} \end{eqnarray}
where $M_\alpha$ denotes the Mn  amplitude $M_\alpha^{\rm M}$, and
$T_1$ and $T_2$ denotes the Tb amplitudes, $M_z^{T,1}$ and
$M_z^{T,2}$, respectively (all for $\Gamma_3$).  In writing this
form we have used the fact that the reality of $f_2^{(3)}$
requires the matrix to be Hermitian.  Also the matrix
elements $a$-$h$ are real, as can be shown from the arguments used
previously for NVO.  We now consider complex-valued matrix
elements $z_n$, which have no analog for NVO.\cite{ANALOG}
We see that the form of Eq. (\ref{MATRIX}) implies a
contribution to $f_2^{(3)}$ of the form $z_1 M_x^* T_1$.  Using inversion
symmetry this term can also be written as $z_1 T_2^* M_x$.  Comparing
this result to that of Eq. (\ref{MATRIX}) we see that
$z_2^*=z_1$.  Similarly one can show that
$z_4^*=z_3$ and $z_6^*=z_5$.  Inversion symmetry gives no
information on the phase of $z_7$.  Thus
the matrix for $f_2^{(3)}$ is of the form
\begin{eqnarray}
\left[ \begin{array} {c c c c c}
a & b & c & y_1 & y_1^* \\
b & d & e & y_2 & y_2^* \\
c & e & f & y_3 & y_3^* \\
y_1^* & y_2^* & y_3^* & g & y_4 \\
y_1 & y_2 & y_3 & y_4^* & h \\
\end{array} \right] \ .
\end{eqnarray}
One can then show that any eigenvector of this matrix must be of the form
\begin{eqnarray}
\psi &=& [ M_x, M_y, M_z, T_1, T_2] =
\sigma_{\rm HTI} [ r_x, r_y, r_z, c, c^* ] e^{i \phi_{\rm HTI}} \ ,
\end{eqnarray}
where $r_\alpha$ is real, $c$ can be complex, and we require the
normalization $\sum r_\alpha^2 + 2 |c|^2 =1$.  As for NVO,
we introduced an arbitrary overall phase $\phi_{\rm HTI}$. Note that
$M_\alpha^{\rm M} (q) = \sigma_{\rm HTI} r_\alpha e^{i \phi_{\rm HTI}}$,
$M_z^{{\rm T},1} (q) = \sigma_{\rm HTI} c e^{i \phi_{\rm HTI}}$ and
$M_z^{{\rm T},2} (q) = \sigma_{\rm HTI} c^* e^{i \phi_{\rm HTI}}$.
Thus, as a result of inversion symmetry, the
amplitudes of the two Tb sublattices are equal in magnitude,
and have equal and opposite relative phases (from $c$ and $c^*$),
the value of which is not fixed by symmetry.  As for NVO,
one can verify that $\psi$ is inversion invariant if
$\phi$ is redefined to be zero, since then
${\cal I} r_\alpha = r_\alpha$ and ${\cal I}c = (c^*)^*=c$.

\subsection{Summary}

\begin{center}
\begin{figure}
\includegraphics[width=12cm]{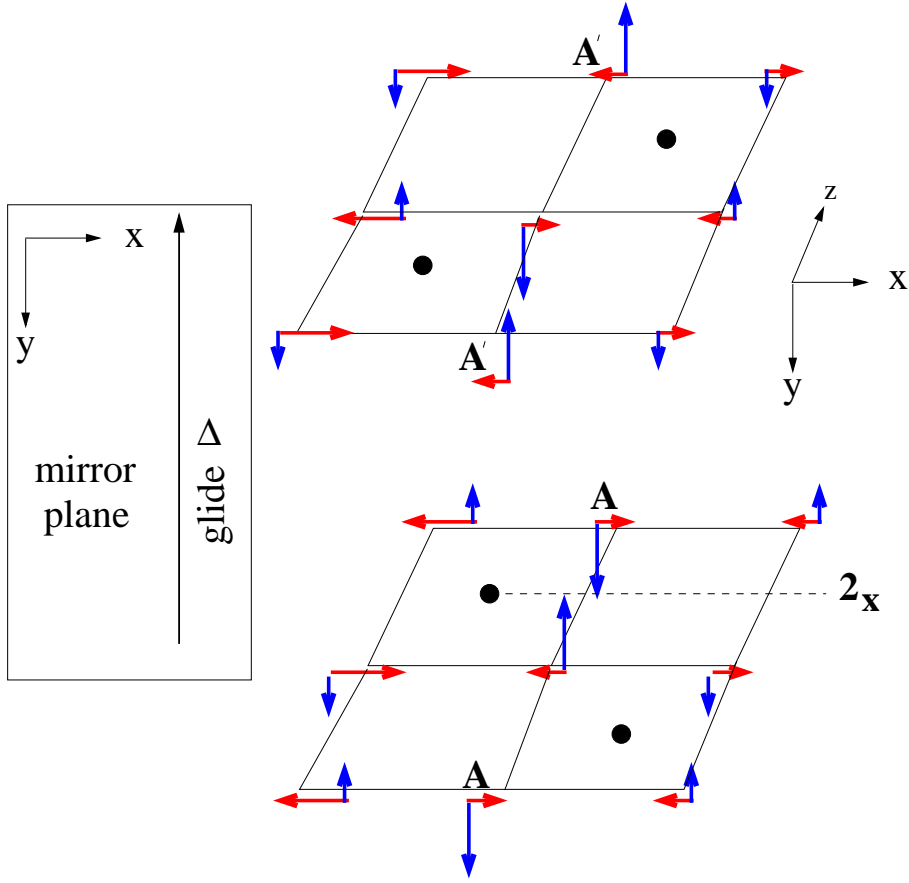}
\caption{Schematic representation to show the symmetry of spin  components
for NVO.  Here we show sections of two adjacent ${\bf a}$-${\bf c}$ planes.
The filled circles indicate the cross-tie sites whose spin
components are not shown, for simplicity. 
The $x$ components transform according $\Gamma_4$: they are odd
under the two-fold rotation $2_x$ about the $x$-axis and are even under the
$x-y$ glide plane (the mirror plane is shown at left with a subsequent
displacement $\vec \Delta$ along $\hat y$).  (Remember that the magnetic
moment vector involves a cross product and therefore is an axial vector:
under a mirror operation it picks up an extra minus sign.)   The $y$
components transform according to $\Gamma_1$: they are even under both
operations. Although the $x$ and $y$ components have different symmetry,
they can plausibly result from nearly isotropic exchange interactions.}
\label{SYM}
\end{figure}
\end{center}

Finally, we should emphasize that although we do not have
a quantitative treatment of the development of magnetic long
range order, we can certainly determine the magnetic symmetry.
This information is encoded in Table \ref{CHAR}.  For
NVO, $\sigma_{\rm HTI}$ is associated with irrep \#4
and therefore is odd under a two-fold rotation about $x$
and even with respect to the mirror plane taking $z$ into $-z$.
Likewise $\sigma_{\rm LTI}$ is associated with irrep \#1
and is therefore even with respect to both these operations.
For future reference we also give the transformation properties of
$\sigmav_{\rm HTI} \sigmav_{\rm LTI}$.  These results are summarized
in Table \ref{SYMTAB}.  The symmetry of the LTI phase of NVO is
illustrated in Fig. \ref{SYM}.

\begin{table}
\caption{Transformation Properties of Order Parameters for NVO
(left) and TMO (right).  In this table $\sigmav \equiv \sigma e^{i \phi}$.
and ``c.c." denotes complex conjugation. Each column gives the result
of applying the operator at the top of the column to the order
parameter listed in the row.}

\vspace{0.1 in}
\begin{tabular} { || c | c | c | c || }
\hline\hline Order Parameter &  $2_x$ & $m_{xy}$ & ${\cal I}$ \\
\hline $\sigmav_{\rm HTI} (q)$ & $-1$ & $1$ & \ c.c.\ \\
$\sigmav_{\rm LTI} (q)$ & $1$ & $1$ & c.c. \\
$\sigmav_{\rm LTI} (q) \sigmav_{\rm HTI} (-q)$
& $-1$ & $1$ & c.c. \\ \hline \hline \end {tabular} \hspace{0.5 in}
\begin{tabular} { || c | c | c | c || }
\hline \hline Order Parameter &  $\tilde 2_y$ & $m_{xy}$ & ${\cal I}$ \\
\hline $\sigmav_{\rm HTI} (q)$ & $-1$ & $1$ & c.c. \\
$\sigmav_{\rm LTI} (q)$ & $1$ & $-1$ & c.c. \\
$\sigmav_{\rm LTI} (q) \sigmav_{\rm HTI} (-q)$
& $-1$ & $-1$ & c.c. \\ \hline \hline \end{tabular}
\label{SYMTAB} \end{table}

\begin{table}
\caption{Values of the symmetry adapted parameters which describe
the HTI and LTI phases of NVO$^{\rm a}$ and TMO.$^{\rm b}$
The uncertainty in the last
digit quoted is given in parenthesis.  Where there is no parenthesis,
the entry is fixed by symmetry to be zero. For TMO the two
Tb order parameters were assumed to have the same magnitude (as
predicted by Landau theory) and the phase difference between the
two Tb parameters in the LTI phase was found to be $1.3(3)\pi$. 
For NVO $T=7$K is in the HTI phase and $T=5$K is in the LTI phase.
For TMO $T=35$K is in the HTI phase and $T=15$K is in the LTI phase.} 

\vspace{0.2 in}
\begin{tabular} {|| c | c || c | c | c ||| c | c || c | c | c ||}
\hline \hline
\multicolumn{5} {|c|} {NVO} & \multicolumn{5} {|c|} {TMO} \\
\hline
\ \ $T$(K)\ \ & Variable & $\alpha =x$ & $\alpha = y$ & $\alpha =z$ &
\ \ $T$(K)\ \ & Variable & $\alpha =x$ & $\alpha = y$ & $\alpha =z$ \\
\hline
7 & $m_{\alpha s}^{(4)}$ & \ \ 1.93(5)\ \ &\ \ 0.20 (5)\ \ &\ \ 0.10 (4)\ \ &
35 & $M^{{\rm M},3}_\alpha $ &\ \ 0.0(8)\ \ &\ \ 2.90(5)\ \ &\ \ 0.0 (5)\ \ \\
7 & $m_{\alpha c}^{(4)}$ & 0 & $-$0.2 (2) & 0.00 (2) &
35 & $M^{{\rm T},3}_\alpha $ & 0 & 0 & 0.0 (4) \\
\hline
5 & $m_{\alpha s}^{(4)}$ & 2.0(1) & 0.16 (9) & 0.01 (5) &
15 & $M^{{\rm M},3}_\alpha $ & 0.0(5) & 3.9(1) & 0.0 (7) \\
5 & $m_{\alpha s}^{(1)}$ & 0.5 (5) & $-$0.5 (1) & 0.00 (3) &
15 & $M^{{\rm M},2}_\alpha $ & 0.0(1) & 0.0(8) & 2.8 (1) \\
5 & $m_{\alpha c}^{(4)}$ & 0 & $-$2.1 (2) & $-$0.03 (9) &
15 & $M^{{\rm T},3}_\alpha $ & 0 & 0 & 0 (1) \\
5 & $m_{\alpha c}^{(1)}$ & 0.9 (5) & 0 & 0 &
15 & $M^{{\rm T},2}_\alpha $ & 1.2(1) & 0(1) & 0 \\
\hline \hline
\end{tabular}

\vspace{0.2 in} \noindent
a) See Ref. \onlinecite{PRL}.
\hspace {2.6 in} b) See Ref. \onlinecite{TMO}.
\label{DATA} \end{table}

For TMO the HTI order parameter $\sigmav_{\rm HTI}$ is odd with respect to the
mirror taking $z$ into $-z$ and is even with respect to the mirror taking
$x$ into $-x$.   Likewise $\sigmav_{\rm LTI}$ is associated with
irrep \#2 and is even with respect to the mirror taking
$z$ into $-z$ and is odd with respect to the mirror taking
$x$ into $-x$ and these results are summarized in Table \ref{SYMTAB}.

In Table \ref{DATA} we give the experimentally determined
values of the symmetry adapted parameters that describe the
HTI and LTI phases of NVO and TMO. The results for NVO are analyzed in
detail in Refs. \onlinecite{NVO1} and \onlinecite{NVO2}. We will 
make a few brief observations here.  For NVO the spine spins dominantly have
order in the $x$-direction in the HTI phase from irrep $\Gamma_4$
indicating that the $x$-axis is the easy axis.   The additional order
in the LTI phase due to irrep $\Gamma_1$ is along the $y$-direction,
as illustrated in Fig. \ref{SYM}.  From this figure one sees that
interactions between nearest neighboring spins in adjacent spines
displaced from one another along either ${\bf c}$ or ${\bf b}$ are
antiferromagnetic.  Since the wavevectors are the same for both
types of order, we infer that the exchange interactions are nearly
isotropic.  For the Mn spins in TMO the situation is much the same.
In the HTI phase, the Mn spins dominantly have order in the $y$ direction,
indicating that this axis is the easy axis.  In the HTI irrep
($\Gamma_3$) one sees, from Table \ref{MNSPIN}, that sites \#1 and \#2
(in one basal plane) have positive $y$-components of spin and that
sites \#3 and \#4 (in the adjacent basal plane) have negative $y$-components
of spin indicating ferromagnetic in-plane interactions and antiferromagnetic
out-of-plane interactions. In the LTI phase of TMO, the additional irrep
$\Gamma_2$ involves spins along $z$-axis and Table \ref{MNSPIN}
shows that for irrep \#2 the components are again arranged ferromagnetically
within basal planes but antiferromagnetically between adjacent basal planes.
The fact that both components of spin
are organized similarly suggests that the exchange interactions are
probably nearly isotropic.

\section{Magnetoelectric Coupling}

\subsection{Landau Theory with Two Order Parameters}

Now we consider the Landau expansion for the free energy, $f_{ME}$,
of the combined magnetic and electric system.  One might be tempted
to write
\begin{eqnarray}
f_{ME} &=& a (T-T_F) {\bf P}^2 + b (T-T_M) |\sigma |^2 \ ,
\end{eqnarray}
where $\sigma$ is a magnetic order parameter and, if we wish to
describe a phase transition in which both electric and magnetic
order appear simultaneously, we would set $T_F=T_M$.  There
are several reasons to reject this scenario.  First of all,
it is never attractive to assume an accidental degeneracy
($T_F=T_M$).  This degeneracy can happen, of course, but
normally one would have to adjust some addition control parameter
(such as pressure) to reach such a higher order critical point.
In addition, in this type of scenario magnetic and electric properties
would not be interrelated.  In NVO and TMO, in contrast, as shown in
Fig, \ref{GL3}, the electric polarization has a dramatic dependence on
the applied magnetic field,\cite{RAPID}  which such an independent
scenario could not explain.

\subsection{Landau Theory with Two Coupled Order Parameters}

Accordingly, we turn to a formulation in which the appearance of magnetic
order induces ferroelectric order.  (The possibility that electric order
induces magnetic order is not allowed by symmetry, by the
argument in footnote 87 of Ref. \onlinecite{toledano}.)  So we write
\begin{eqnarray}
f_{ME} &=& a \chi_E^{-1} {\bf P}^2 + a (T-T_M) |\sigma |^2 + V_{ME} \ ,
\end{eqnarray}
where $\chi_E^{-1}$ does not approach zero and the simultaneous
appearance of magnetic and electric order is due to the term $V_{ME}$.
As we have seen, the magnetic order is associated with a nonzero wavevector,
whereas the ferroelectric order is a zero  wavevector phenomenon.
Accordingly, we are constrained to posit a magnetoelectric coupling
of the form
\begin{eqnarray}
V_{ME} & \sim & \sigma(q) \sigma(-q) P \ .
\end{eqnarray}
This term will do what we want: when magnetic order appears
in $\sigma (q)$, it will then give rise to a linear perturbation
in $P$, so that $P \sim\chi_E |\sigma (q)|^2$.  This argument is
schematic, of course, and we will have to fill in the details,
which must be consistent with the crystal symmetry of the
specific systems involved. 

The minimal phenomenological model which describes the magnetic and electric
behavior of the HTI and LTI phases is therefore written as
\begin{eqnarray}
f &=& \oh (T-T_{\rm HTI}) \sigmav_{\rm HTI} (q) \sigmav_{\rm HTI}^*(q)
+ \oh (T-T_{\rm LTI}) \sigmav_{\rm LTI} (q) \sigmav_{\rm LTI}^*(q)
\nonumber \\ && \ \ 
+ {\cal O} (|\sigma|^4) + \oh \chi_E^{-1} {\bf P}^2 + V_{ME} \ ,
\end{eqnarray}
where
\begin{eqnarray}
V_{ME} &=& \sum_{\rm A,B = LTI, HTI} \sum_{\gamma=x,y,z}
a_{A,B,\gamma} \sigmav_A (q)^* \sigmav_B (q) P_\gamma \ .
\end{eqnarray}

\subsection{Symmetry of Magnetoelectric Coupling}

We now show that this free energy reproduces the observed phenomenology
of ferroelectricity in NVO and TMO. First, of all, in the HTI phase
(where $\sigmav_{\rm LTI}=0$) $V_{ME}$ is of the form
\begin{eqnarray}
V_{ME} &=& \sum_\gamma b_\gamma |\sigmav_{\rm HTI} (q)|^2 P_\gamma \ ,
\label{MEEQ} \end{eqnarray}
where $b_\gamma$ is real.
Now we use the fact that $V_{ME}$ has to be inversion-invariant, since
it arises in an expansion relative to the paramagnetic phase,
which is inversion-invariant.\cite{PSYM1,PSYM2}  We use
${\cal I}\sigma_{\rm HTI}(q)=\sigma_{\rm HTI}(q)^*$ and
${\cal I} P_\gamma = - P_\gamma$ to show that $b_\gamma$ must vanish.
Indeed, we have already seen, the HTI phases of NVO and TMO are inversion
invariant.  So for these situations $b_\gamma$ in Eq. (\ref{MEEQ})
must be zero and no  polarization  can be induced in the HTI phase.

Now we consider the situation in the LTI phase when the two
order parameters $\sigma_{\rm HTI}$ and $\sigma_{\rm LTI}$ are
both nonzero.  The argument which indicated that
$a_{{\rm HTI},{\rm HTI}, \gamma}=0$ can be used to establish that
$a_{{\rm LTI},{\rm LTI}, \gamma}=0$.  Then we write
\begin{eqnarray}
V_{ME} &=& \sum_\gamma [ c_\gamma \sigmav_{\rm HTI} (q)^*
\sigmav_{\rm LTI}(q) + c_\gamma^* \sigmav_{\rm LTI}^*(q) \sigmav_{\rm HTI}(q)
] P_\gamma \ .
\label{PEQ} \end{eqnarray}
This interaction has to be inversion invariant, so we use
the transformation properties of the order parameters under
inversion to write
\begin{eqnarray}
V_{ME} &=& {\cal I} V_{ME} =  - \sum_\gamma [ c_\gamma \sigmav_{\rm HTI} (q)
\sigmav_{\rm LTI}(q)^* + c_\gamma^* \sigmav_{\rm LTI}(q)
\sigmav_{\rm HTI}(q)^* ] P_\gamma \ .
\end{eqnarray}
Comparison with Eq. (\ref{PEQ}) indicates that $c_\gamma$ must be pure
imaginary: $c_\gamma=ir_\gamma$, where $r_\gamma$ is real.  Then
\begin{eqnarray}
V_{ME} &=&   2 \sum_\gamma r_\gamma \sigma_{\rm HTI} (q)
\sigma_{\rm LTI}(q) P_\gamma \sin [ \phi_{\rm HTI} - \phi_{\rm LTI} ] \ .
\label{TEMP} \end{eqnarray}
This result shows that to get a nonzero spontaneous polarization it
is necessary that two order parameters be nonzero.
(A similar interaction was proposed by Frohlich {\it et al.}\cite{TWO}
in their analysis of second harmonic generation.) Furthermore, these
two order parameters must {\it not} have the same phase.
In fact, a more detailed analysis of Landau theory shows that the
phase difference $\phi_{\rm HTI} - \phi_{\rm LTI}$ is expected to
be $\pi /2$.  (This result comes from an analysis of the quartic
terms.  As we observed earlier, the function of the quartic terms
is to enforce the constraint of fixed spin length.  This constraint
usually means that the ordering in two representations should be
out of phase, so that when one representation gives a maximum of
spin  lengths, the other gives a minimum of spin lengths.)

\begin{center}
\begin{figure}
\includegraphics[width=12cm]{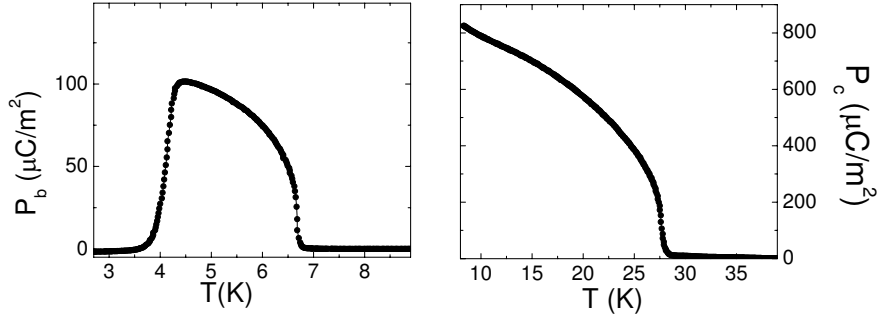}
\caption{Adapted from Refs. \protect{\onlinecite{RAPID}} and
\protect{\onlinecite{KIM}}.
Temperature dependence of the spontaneous polarization at
zero applied magnetic field for NVO (left) and for TMO (right).
Note the different scales for ${\bf P}$.}
\label{POLOFT}
\end{figure}
\end{center}

Finally, we consider how the symmetry properties constrain the
spontaneous polarization.  Look at Table \ref{SYMTAB}.  There
we see how the magnetic order parameters transform under
the various symmetry operations of the paramagnetic phase.
For $V_{ME}$ to be an invariant, we see that for NVO, ${\bf P}$
must be odd under $2_x$ (which restricts ${\bf P}$ to be along
$y$ or $z$) and it must be even under $m_{xy}$ (which restricts
${\bf P}$ to be along $x$ or $y$).  Thus, symmetry restricts ${\bf P}$
to be only along $y$.  This is exactly what experiment shows.
For TMO, ${\bf P}$ must be odd under both $\tilde 2_y$ and $m_{xy}$.
Thus, symmetry restricts ${\bf P}$ to lie along $z$ at $H=0$, as
is observed in experiment. (At higher magnetic fields the magnetic symmetry
must change to  explain why the polarization switches from the
$z$-axis to the $x$-axis.)  Furthermore, the temperature variation
of ${\bf P}$, shown in  Fig.  \ref{POLOFT}
looks very much like that for an order parameter.  But that is to  be
expected because if we minimize the total free energy with
respect to ${\bf P}$, using Eq. (\ref{TEMP}), we see that the
spontaneous polarization is given as
\begin{eqnarray}
P_\gamma \sim \chi_E \sigma_{\rm HTI} \sigma_{\rm LTI} \ .
\label{PSSEQ} \end{eqnarray}
When the LTI phase is entered, $\sigma_{\rm HTI}$ is already
well developed and is therefore essentially independent of
temperature.  Thus we expect that crudely $P_\gamma \sim \sigma_{\rm LTI}$.
Indeed, although we have not undertaken a quantitative analysis,
the experimental curves of $P$ versus $T$ look quite similar to
those for an order parameter.

Finally, for TMO for a large magnetic field along ${\bf a}$ (see Fig.
\ref{PHASE}) or along ${\bf b}$ (see Ref. \onlinecite{TKPRB}),
there is a change of orientation of
the spontaneous polarization ${\bf P}$ to lie along ${\bf a}$. Since
there seems to be no analogous phase transition within the HTI phase,
we attribute this reorientation to a change in the LTI spin state.
Instead of the additional irrep of the LTI phase being $\Gamma_2$
(as it is at low field), we infer that the new LTI irrep is $\Gamma_1$,
since this combination of irreps is consistent with having ${\bf P}$
along ${\bf a}$. Furthermore, if we assume that the exchange coupling
is isotropic, then we would expect that $\Gamma_1$ ordering would
be ferromagnetic within basal planes and antiferromagnetic between planes.
From Table \ref{MNSPIN} this constraint can only be satisfied if the
ordering involves the $x$-component of spin.  So, from the polarization
data we speculate that the Mn spin structure (which at low field is in the 
{\bf b}-{\bf c} plane) is rotated, at high field,
into the ${\bf a}$-${\bf b}$ plane.

\subsection{Broken Symmetry}

We should also mention some considerations concerning broken symmetry
for NVO. (Clearly a similar discussion applies to other similar systems.)
Since both transitions involving the HTI phase involve broken
symmetry we assert the following.  At the level of the present
analysis when the temperature is reduced to enter the HTI phase,
the modulated order appears with an arbitrary phase $\phi_{\rm HTI}$.
Of course, if this state is truly incommensurate, then the phase
will remain arbitrary.  Normally, however, we would expect some
perturbation to break this symmetry and this continuous symmetry
should be removed.  However, we do expect a degeneracy with respect
to the time-reversed version of the ordered HTI phase.  In that case
upon performing many runs of the same experiment, both versions
of the HTI phase should occur with equal probability.  

One can make much the same observation about the HTI$\rightarrow$LTI phase
transition.  Here one has the additional broken symmetry associated with
the irrep $\Gamma_1$.  When the temperature is reduced to enter the
LTI phase, the system will have two symmetry-equivalent states into
which it can condense.  As with the usual magnetic phase transitions,
one can (in principle) select between these two phases by applying a
suitably spatially modulated magnetic field.  Such an experiment
does not seem currently feasible (because modulation of an applied
field on an atomic scale is difficult to produce).  However, because
the magnetic order parameters are coupled to the ferroelectric moment,
one can select between the two symmetry equivalent possibilities for
the LTI order parameter by applying a small {\it electric} field. 
A interesting experiment suggests itself: compare the magnetic
state as determined by, say, neutron diffraction for the two
cases of a small applied electric field in the positive and
negative $\underline b $ directions.  According to the
magnetoelectric trilinear coupling, application of such an
electric field should select the sign of the product
$\sigma_{\rm HTI} \sigma_{\rm LTI}$.  In this context we remark
that measurement of the spontaneous polarization ${\bf P}$ (as
in Fig. \ref{GL3}) is made by preparing the sample in a small 
symmetry-breaking electric field ${\bf E}_0$, which is removed
once ${\bf P}$ becomes nonzero.  The ferroelectric order is confirmed
by verifying that ${\bf P}$
changes sign when the sign of ${\bf E}_0$ is changed. 

\section{MICROSCOPICS}

\begin{center}
\begin{figure}
\includegraphics[width=12cm]{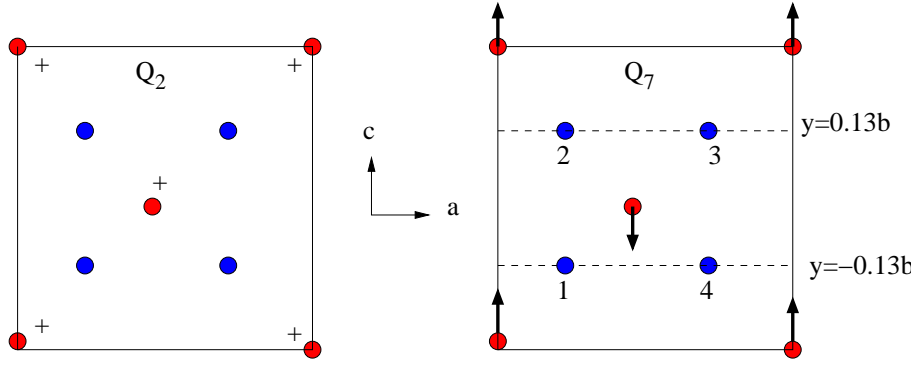}
\caption{Pattern of two generalized displacements, $Q_2$ (left) and
$Q_7$ (right), which transform under the symmetry operations of the
crystal like a displacement along ${\bf b}$.
In the left panel "$+$" indicates motion along the positive {\bf b} axis.
In $Q_2$ all the cross-tie sites move in parallel along the ${\bf b}$ axis
and therefore this motion induces a dipole moment. As discussed in the
text, the nonuniform motion of the cross-tie sites in the generalized
displacement $Q_7$ induces uniform motion of the spine site in the ${\bf b}$ 
direction which in turn produces a dipole moment.}
\label{PHONON}
\end{figure}
\end{center}

Since the spontaneous  polarization ${\bf P}$ must result from a spontaneous
condensation of an optical phonon having a dipole moment, we are led to
study the symmetry of the phonon excitations at zero wavevector.  Neglecting
nonzero wavevectors, we write the $\alpha$-component of the displacement of
the $\tau$th ion in the unit cell at ${\bf R}$ as
\begin{eqnarray}
u_\alpha ({\bf R},\tau) &=& \sum_i Q_i \xi^{(i)}_\alpha (\tau) \ ,
\end{eqnarray}
where $\xi^{(i)}_\alpha (\tau)$ is the normalized form factor of the $i$th
generalized displacement whose amplitude is $Q_i$. A
comprehensive analysis is given elsewhere,\cite{ELSE} but here we
confine our attention to generalized displacements in NVO which transform
appropriately (like a displacement along {\bf b}) to explain
experiments.  Such a generalized displacement $Q_i$ 
must be invariant under the operations
(see Table II) $E$, $\sigmav_x$, $\sigmav_z$ and $2_y$ and change
sign under $\sigmav_y$, $2_x$, $2_z$, and ${\cal I}$.  There are 12 such
generalized displacements of the 13 ions in the primative unit cell.  Six of
these are the uniform displacements along ${\bf b}$ of all
crystallographically equivalent sites of a given type, {\it viz.} Ni spine sites,
Ni cross-tie sites, V sites, O$_1$, O$_2$, and O$_3$ sites, and
these uniform displacements, denoted $Q_1$, $Q_2$ ... $Q_6$,
give rise to a dipole moment along the ${\bf b}$ axis.  Other
generalized displacements involve, perhaps surprisingly, oppositely oriented
displacements along the ${\bf a}$ or ${\bf c}$ axis within a group of
crystallographically equivalent sites.  We illustrate one of these
($Q_7$ involving Ni cross-tie sites) along with $Q_2$ in Fig. \ref{PHONON}.
Since $Q_7$ has the same symmetry as $Q_1$ ... $Q_6$, it must couple
to these modes.  One can easily visualize this by imagining the ions
to act like hard spheres.  In that case, as the cross-tie ions
approach spine sites \#1 and \#4, they cause these site (which initially
were at negative $y$) to move to more negative $y$.  Similarly, as 
the cross-tie sites move away from sites \#2 and \#3, these ions have
more room to move closer to $y=0$.  In other words, the opposing motion of the
cross-tie sites in mode $Q_7$ along the ${\bf c}$ axis interacts with
the uniform motion in mode $Q_1$ of the spine sites along ${\bf b}$. 
In summary, the elastic free energy as a function of displacements can
be written as

\begin{eqnarray}
f(\{Q_i\}) &=& {1 \over 2} \sum_{i,j=1}^{12} V_{ij} Q_i Q_j \ .
\end{eqnarray}
At the time of this writing no calculation or neutron experiments to 
determine $V_{ij}$ have appeared.  Instead we have recourse to a very
crude toy model, obtained by setting
\begin{eqnarray}
f(\{Q_i\}) &=& \sum_i {1 \over 2} M_i \omega_{\rm D}^2 Q_i^2 \ ,
\label{AA} \end{eqnarray}
where $M_i$ is the mass of ions in mode $Q_i$ and $\omega_{\rm D}$
is the Debye frequency, characteristic of phonons.

We now consider the effect of a generalized displacement $Q_i$ on the
exchange interaction between nearest neighbors in the same spine.
Then for spins numbered 1 and 4 in a unit cell we have the exchange
interaction as a function of displacement as
\begin{eqnarray}
{\cal H}_{14} (Q_i) &=& {\cal H}_{14}(0) + Q_i \sum_{\alpha \beta}
S_\alpha (1) {\partial M_{\alpha \beta} (1,4) \over \partial Q_i}
S_\beta (4) \ .
\end{eqnarray}
The existence of a mirror plane perpendicularly bisecting the 1-4 bond
($m_{\bf bc}$) implies that
\begin{eqnarray}
{\cal H}_{14}(Q_i) &=& m_{\bf bc} {\cal H}_{14}(Q_i) \ ,
\end{eqnarray}
which is
\begin{eqnarray}
{\cal H}_{14}(Q_i) &=& Q_i {d \over dQ_i} 
[ S_{1x}, S_{1y}, S_{1z}] {d \over dQ_i} \left[
\begin{array} {c c c} M_{xx} & -M_{yx} & -M_{zx} \\ -M_{xy} & M_{yy} &
M_{zy} \\ -M_{xz} & M_{yz} & M_{zz} \\ \end{array} \right] \left[
\begin{array} {c}  S_{4x} \\ S_{4y} \\ S_{4z} \\ \end{array} \right] \ ,
\end{eqnarray}
where we used $m_{\bf bc}Q_i=Q_i$, $m_{\bf bc}S_{1x}=S_{4x}$,
$m_{\bf bc}S_{1y}=-S_{4y}$, and $m_{\bf bc}S_{1z}=-S_{4z}$
(the spin is a pseudovector).  Thus the derivative must have the form
\begin{eqnarray}
{\partial M_{\alpha \beta} (1,4) \over \partial Q_i} &=& {\partial \over
\partial Q_i} \left[ \begin{array} {c c c} J_{xx} & D_z & - D_y \\
-D_z & J_{yy} & J_{yz} \\ D_y & J_{yz} & J_{zz} \\ \end{array}
\right] \ ,
\label{CC} \end{eqnarray}
where $J_{\alpha \beta}$ is the symmetric exchange tensor and ${\bf D}$ 
is the Dzialoshinskii-Moriya vector, which specifies the antisymmetric 
component of the exchange tensor.

We determine the other similar
interactions in the unit cell using the appropriate symmetry operations.
If $2_y$ is a rotation about an axis parallel to ${\bf b}$ and which
passes through site \#4, then
\begin{eqnarray}
{\cal H}(4,1';y) &=& 2_y {\cal  H} (1,4;y) \nonumber \\ &=&
Q_i [S_{4x}, S_{4y} , S_{4z}] {d \over dQ_i} \left[
\begin{array} {c c c} J_{xx} & D_z & D_y \\ -D_z & J_{yy} &
-J_{yz} \\ -D_y & -J_{yz} & J_{zz} \\ \end{array} \right] \left[
\begin{array} {c}  S_{1'x} \\ S_{1'y} \\ S_{1'z} \\ \end{array} \right] \ ,
\end{eqnarray}
where we used $2_y Q_i = Q_i$, and site \#4' is one unit cell to the
right of site \#4 in Fig. \ref{PHONON}.  Also, if $2_x$ is a rotation
about the ${\bf a}$ axis, then
\begin{eqnarray}
{\cal H}(2,3;y) &=& 2_x {\cal H}(1,4;y) \nonumber \\ &=&
Q_i [S_{2x}, S_{2y} , S_{2z}] {d \over dQ_i}
\left[ \begin{array} {c c c} -J_{xx} & D_z & -D_y \\ -D_z & -J_{yy} &
-J_{yz} \\ D_y & -J_{yz} & -J_{zz} \\ \end{array} \right]
\left[ \begin{array} {c}  S_{3x} \\ S_{3y} \\ S_{3z} \\ \end{array}
\right] \ ,
\end{eqnarray}
where we used $2_xQ_i =-Q_i$ and
\begin{eqnarray}
{\cal H}(3,2';y) &=& 2_y {\cal H}(2,3;y) \nonumber \\
&=& Q_i [S_{3x}, S_{3y} , S_{3z}] {d \over dQ_i}
\left[ \begin{array} {c c c} -J_{xx} & D_z & D_y \\ -D_z & -J_{yy} &
J_{yz} \\ -D_y & J_{yz} & -J_{zz} \\ \end{array} \right] \left[
\begin{array} {c}  S_{2'x} \\ S_{2'y} \\ S_{2'z} \\ \end{array} \right] \ ,
\end{eqnarray}
where site \#2' is one unit cell to the right of site \#3 in Fig. \ref{PHONON}

When we consider Eq. \ref{AA}) and neglect fluctuations, the spin-phonon
interactions lead to the result
\begin{eqnarray}
\langle Q_i \rangle &=& (M_i \omega_{\rm D}^2)^{-1} \sum_{\alpha \beta}
\sum_{n,m} \langle S_\alpha (n) \rangle {\partial M_{\alpha \beta} (n,m)
\over \partial Q_i} \langle S_\beta (m) \rangle \ ,
\label{BB} \end{eqnarray}
where $\langle \ \ \rangle$ indicates a thermal average and $(n,m)$ are
summed over the 4 nearest-neighbor spine-spine interactions in a unit cell.  
Assuming the spins are characterized by spine spin components
scaled by ${\bf a}$ for irrep 4 and by ${\bf b}$ for irrep \#1, we
write the spin components as
\begin{eqnarray}
S_x (x_1) &=& (a_x+ib_x) e^{iqx_1} + (a_x^*-ib_x^*) e^{-iqx_1} \nonumber \\
S_x (x_2) &=& (-a_x+ib_x) e^{iqx_1} + (-a_x^*-ib_x^*) e^{-iqx_1} \nonumber \\
S_x (x_3) &=& (a_x-ib_x) e^{iqx_4} + (a_x^*+ib_x^*) e^{-iqx_4} \nonumber \\
S_x (x_4) &=& (-a_x-ib_x) e^{iqx_4} + (-a_x^*+ib_x^*) e^{-iqx_4}
\end{eqnarray}
\begin{eqnarray}
S_y (x_1) &=& (ia_y+b_y) e^{iqx_1} + (-ia_y^*+b_y^*) e^{-iqx_1} \nonumber \\
S_y (x_2) &=& (ia_y-b_y) e^{iqx_1} + (-ia_y^*-b_y^*) e^{-iqx_1} \nonumber \\
S_y (x_3) &=& (-ia_y+b_y) e^{iqx_4} + (ia_y^*+b_y^*) e^{-iqx_4} \nonumber \\
S_y (x_4) &=& (-ia_y-b_y) e^{iqx_4} + (ia_y^*-b_y^*) e^{-iqx_4} 
\end{eqnarray}
\begin{eqnarray}
S_z (x_1) &=& (a_z+ib_z) e^{iqx_1} + (a_z^*-ib_z^*) e^{-iqx_1} \nonumber \\
S_z (x_2) &=& (a_z-ib_z) e^{iqx_1} + (a_z^*+ib_z^*) e^{-iqx_1} \nonumber \\
S_z (x_3) &=& (a_z-ib_z) e^{iqx_4} + (a_z^*+ib_z^*) e^{-iqx_4} \nonumber \\
S_z (x_4) &=& (a_z+ib_z) e^{iqx_4} + (a_z^*-ib_z^*) e^{-iqx_4} \ .
\end{eqnarray}

Using these evaluations one can carry out the sum over $(n,m)$ in
Eq. (\ref{BB}) to get
\begin{eqnarray}
\langle Q_i \rangle = 16 (M_i \omega_{\rm D}^2)^{-1} [F_i^{(s)} \sin (qa/2)
+ F_i^{(c)} \cos(qa/2)] \ ,
\end{eqnarray}
where
\begin{eqnarray}
F_i^{(c)} &=& \Im [a_x^* b_z + a_z^* b_x] d D_y /dQ_i
+ \sum_\alpha \pi_\alpha \Im [a_\alpha b_\alpha^* ] dJ_{\alpha \alpha} /dQ_i 
\end{eqnarray}
and
\begin{eqnarray}
F_i^{(s)} &=& \Im [a_zb_y^* + b_za_y^*]d J_{yz} /d Q_i +
\Im [a_xb_y^* + b_xa_y^*]d D_z /d Q_i \ ,
\end{eqnarray}
where $-\pi_x=\pi_y=\pi_z=1$.
Note that these terms require the presence of both order parameters ${\bf a}$
and ${\bf b}$ and hence they can only be nonzero in the LTI phase. Also
these terms are only nonzero if ${\bf a}$ and ${\bf b}$ have different phases.
For displacements which could give rise to a spontaneous polarization along the
${\bf a}$ or ${\bf c}$ axes, the sum over $(n,m)$ in Eq. (\ref{BB}) gives
zero.\cite{ELSE}
These conclusions agree with the result found using Landau theory.
This magnon-phonon coupling also contributes
to the temperature dependence of the wavevector $q$.\cite{NVO1,NVO2}

To get an order-of-magnitude estimate of the various quantities,
we consider the effect of the motion of the oxygen ions, which are
the lightest atoms and therefore have the largest displacements.
Crudely speaking, the dipole moment, $P_Q$ of the generalized displacement
$Q$ is given by $P_Q= q_Q \sum_{i \in Q} u_i$, where $q_Q$ is the charge
of the ions of $Q$ and $u_i$ is the displacement of ion $i$ in $Q$.
More accurately, $q_Q$ should be replaced by an effective charge
$q^*_Q$ which takes account of the electrical relaxation which occurs
as the ions move.  (This is analogous to the discussion given at the
end of the preceding paragraph.)  Thus even $Q_7$ will develop a
(probably small) dipole moment in the ${\bf b}$ direction.
However, for simplicity we set
\begin{eqnarray}
\langle P \rangle = q_i \langle Q \rangle / v_{\rm uc} = 2e
\langle Q \rangle / v_{\rm uc} \ ,
\end{eqnarray}
where $v_{\rm uc} \approx 275 \times 10^{-30}$m$^3$ is the volume of
the unit cell.  More accurately,
$\langle Q \rangle$ should be replaced by $\langle Q \rangle \sqrt n$,
where $n$ is the number of ions involved in the mode generalized
displacement $Q$. (So $n=4$ or $n=6$.) So, in meters,
$\langle Q \rangle \approx (275 \times 10^{-30})
(5 \times 10^{-4}) /(3.2 \times 10^{-19}\sqrt n)$, where we took
$P=5 \times 10^{-4}$C/m$^2$ as a typical value.   Thus we estimate
the ionic displacement to be of order $\langle Q \rangle \sim 0.001\AA$.
(Actually, neutron diffraction indicates that the ionic displacement ought to 
be at most 0.001$\AA$.\cite{CB})  If $\partial J/\partial Q$ represents a
typical value for $\partial M_{\alpha \beta}/dQ$, then
\begin{eqnarray}
\langle Q_i \rangle \sim {(\hbar c)^2 \over (M_ic^2) (\hbar \omega_{\rm D})^2}
{\partial J \over \partial Q}\ .
\end{eqnarray}
Working in $\AA$ and eV and taking $\langle Q_i \rangle = 0.001\AA$,
$\hbar \omega_{\rm D} \approx 0.05$eV,
$M_i c^2 \approx 1.6 \times 10^{10}$eV, and $\hbar c \approx 2000$eV$\AA$,
we find that this mechanism requires that 
\begin{eqnarray}
{\partial J \over \partial Q} \sim 0.01 {\rm eV}/\AA \ .
\end{eqnarray}
This seems to be a plausible value.
Obviously a first principles calculation of $\partial M_{\alpha \beta}
/\partial Q_i$ would be of interest to make this analysis more concrete.

\section{SUMMARY AND OUTLOOK}

The development of multiferroic materials having very large 
magnetoelectric couplings offers the possibility of designing new types of 
devices which exploit the coupling between magnetic and ferroelectric 
order.  Furthermore, investigating the nature of the coupling between 
magnetic and ferroelectric order parameters in these compounds may be 
important in understanding other systems displaying significant 
interactions between different types of long range order.  We will briefly 
summarize the main results of the model we have presented coupling 
ferroelectricity with incommensurate magnetic order, and then discuss what 
this implies for future research on magnetoelectric multiferroics.

\subsection{Summary of this Review}

As many of the recently identified materials exhibiting simultaneous 
magnetic and ferroelectric order are incommensurate magnets, we have 
focused on these systems.  We discussed a toy model for incommensurate 
magnetism.  In this model, we saw
that, under the assumption that the magnetic anisotropy is not too large,
the magnetic system will undergo a paramagnetic to a longitudinally ordered 
incommensurate phase we refer to as the HTI phase.  On further lowering the 
temperature, the is another transition to a distinct incommensurate phase 
with additional transverse spin ordering, which we call the LTI phase.

Because knowing the detailed 
symmetry of the incommensurate magnetic structure is crucial for 
determining whether or not ferroelectric order is allowed, we considered the 
extension of this toy model to systems with non-trivial unit cells.  We 
addressed this problem by expressing the spin order parameters in terms of 
irreducible representations consistent with the symmetry restrictions of 
the unit cell.  The central observation for understanding the 
magnetoelectric coupling is that the free energy must be invariant under 
all symmetries of the paramagnetic phase, and in particular, if the
paramagnetic crystal has inversion symmetry, it must be 
invariant under spatial inversion.  This requirement was used to determine 
whether or not a particular incommensurate magnetic structure allowed the 
possibility of ferroelectric order.  Using this approach, we are able to 
qualitatively explain the multiferroic behavior of both NVO and TMO, 
including the absence of ferroelectric order in the HTI phase, the 
development of ferroelectricity in the LTI phase, and the qualitative 
features of the spontaneous polarization (direction and temperature 
dependence).  

It is worth noting that the magnetoelectric coupling we have
described here does not reduce to the analogous coupling which
can occur in a ferromagnet or in an antiferromagnet.  As remarked
in the review of Smolenskii and Chupis,\cite{REV1} such a trilinear
coupling can not exist in structures which (like NVO or TMO)
have inversion symmetry in the paramagnetic phase. Indeed,
the mechanism we invoke requires that $q \not= 0$, as one can
see from Eq. (\ref{TEMP}).  (If $q=0$, then the order parameters are
real, $\phi_{\rm HTI}=\phi_{\rm LTI}=0$ and $V_{\rm ME}=0$.)
In that review they also mention a coupling which involves
gradients of the magnetic order parameter.  That type of
coupling may be related to that used here, although in our
case the symmetry properties of the unit cell play a crucial role
which can not be replaced by a continuum vector field.

We also showed that the microscopic symmetry of the derivative of the
exchange tensor with respect to ionic displacement leads to results
in complete agreement with the symmetry arguments based on the Landau
expansion.  This symmetry will have to be respected by any truly
microscopic theory of magnetoferroelectrics.

\subsection{Outlook for Device Applications}

The success of the theory described in this review suggests that 
it may be valuable both in understanding the origins of multiferroic
behavior in presently identified systems, and well as in guiding 
the search for new multiferroic compounds having desirable materials 
properties.  We briefly discussed the technological drivers motivating the 
search for magnetoelectric materials, by illustrating the types of devices 
that might be possible using multiferroics.  However, one of the 
difficulties that must be resolved before these systems could possibly be 
incorporated into fabricating next generation magnetoelectric devices is 
that the very low transition temperatures into the ferroelectric LTI 
phase (6.4 K for NVO and $\sim 27$ K for TMO) make these materials 
unsuitable for many applications.  We discuss in the following some of the 
general ideas extracted from our model which may help guide the search for 
new multiferroics having higher transition temperatures and larger 
magnetoelectric couplings.

Extending the search for multiferroics from simple ferromagnets to systems 
with incommensurate magnetic order is an important first step in finding 
materials which have a room temperature transition into a phase with 
simultaneously appearing magnetic and ferroelectric order.  Insulating 
ferromagnets tend to have very low transition temperatures, but many 
incommensurate magnets have ordering temperatures well above room 
temperature.\cite{ABOVE}  In fact, the incommensurate magnetic
structure associated with ferroelectric order in one recently identified
multiferroic\cite{KIMURAFERRITE} persists up to $T=320$ K.  Our results
suggest that insulating incommensurate magnets with high magnetic
ordering temperatures may be prime candidates in the search for
strongly coupled magnetoelectric multiferroics at room temperature.

In the next subsection we propose experiments which might
show that in systems such as those studied here magnetic
phase boundaries can be sensitive to the applied
{\it electric} field.

\subsection{Experimental Outlook}

It is suggestive that for the two systems we have considered in detail,
the development of a single order parameter can {\it not} induce
a spontaneous polarization. A requirement for such a result to be
general is that in the paramagnetic phase the system should have
a center of inversion symmetry.  In that case, it is assured that
in the paramagnetic phase there is zero spontaneous polarization.
We speculate that for such a system, the development of 
incommensurate magnetic order via a continuous phase transition
does not allow a spontaneous polarization to be induced by the
magnetic ordering.  That is, the development of a single magnetic order 
parameter transforming under one specific irrep cannot break the spatial 
inversion symmetry and lead to ferroelectric order.  In order for 
incommensurate magnetic order to break the spatial inversion symmetry (and 
potentially induce the development of ferroelectricity) we postulate that 
it is necessary to have two distinct magnetic order parameters.
(A similar proposal in connection with second harmonic generation has
been made by Frohlich {\it et al.}\cite{TWO})

This speculation raises the more general
question concerning a system which in the disordered phase has neither
magnetic nor ferroelectric order. In particular, consider
systems which lack inversion symmetry, but whose rotational
symmetry elements preclude a nonzero vector order parameter.
We give two families of such crystal structures.  The first is
that of the point group $D_2$ (orthorhombic space groups \#16-\#24 in Ref.
\onlinecite{HAHN}) and the second is that of point group $T$
(cubic space groups \#195 - \#199 in Ref. \onlinecite{HAHN}).
In the para phase, these systems have no magnetic long-range order
and, because these crystal structures do not allow vector ordering,
they do not display
ferroelectric order.  When such a system develops long-range
incommensurate order with a wavevector along one of the
crystallographic directions, then only rotations about
this direction remain symmetries and a spontaneous polarization
along the direction of the wavevector is permitted, at least
in principle.  So in this case, it would seem that two order
parameters would not be necessary for magnetic long range
order to induce ferroelectric order. Accordingly we suggest
that it would be interesting to find ferroelectric incommensurate
magnets having one of these crystal structures. 

\begin{center}
\begin{figure}
\includegraphics[height=5cm]{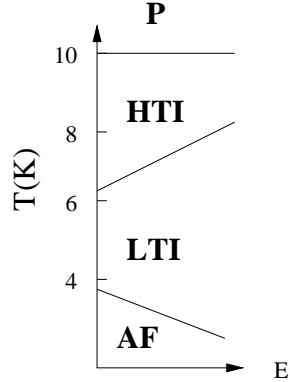}
\caption{Schematic phase diagram for A system like NVO in
the $T$-$E$ plane.  (We do not give the scale of the $E$-axis.)}
\label{EPHASE}
\end{figure}
\end{center}

Another experimental program which this study suggests concerns
the phase diagram of these systems in the $T$-$E$ plane, where
$E$ is the uniform applied field.  Since only the LTI phase of
NVO or TMO has a spontaneous polarization, this phase is favored
(relative to the HTI or AF phases) in the presence of an electric
field.  So we propose the schematic phase diagram shown in
Fig. \ref{EPHASE}.  We have not specified the scale of the
horizontal axis in this schematic figure, but at least for
the LTI-AF transition in NVO we can estimate how thin a 
film would have to be to produce a 5\% shift in the transition
temperature for an applied voltage of 5 V.
The analog of the Clausius-Clapeyron equation for the
LTI-AF phase boundary in the $T$-$E$ plane is
\begin{eqnarray}
dT/dE = - V(P_{\rm LTI}-P_{\rm AF})/(S_{\rm LTI}-S_{\rm AF}) \ .
\end{eqnarray}
Now take $V$ to be the volume per Ni ion.  The volume
of the conventional unit cell is $v=abc$, so $V=abc/12$,
because there are 12 Ni's per conventional unit cell.
$a=5.9$, $b=11.4$, and $c=8.2$, all in $10^{-10}$m.
For one Ni ion the total entropy change from zero to infinite
temperature is $k \ln 3$.  Guided by specific heat
measurements\cite{PRL} we set
\begin{eqnarray}
(S_+-S_-) = 0.01 k \ln 3
\end{eqnarray}
and take $P=5 \times 10^{-4}$C/m$^2$ as a typical value.  Then we find
\begin{eqnarray}
dT/dE &=& 
- [( 540 \times 10^{-30} {\rm m}^3 /12]) (5 \times 10^{-4} {\rm C/m}^2) 
/ [1.4 \times 10^{-25} {\rm J/K}] 
\nonumber \\ &= & 1.5 \times 10^{-7} {\rm Km/V} \ . 
\end{eqnarray}
To see roughly what this means, set $dT=0.05T_c=0.2$K and
$dE = 5{\rm  V}/t$, where $t$ is the thickness of the sample.
This gives $0.2t \times 10^{-8}$, or $t\approx 1 \mu$.

Finally, we emphasize that it would be desirable to determine the
magnetic structure of TMO for high magnetic fields along the
{\bf a} or {\bf b} direction to test whether the magnetic
structure proposed below Eq. (\ref{PSSEQ}) is realized. 

\subsection{Theoretical Outlook}

It is clear that the next step for theorists is to construct a
fully microscopic theory to explain the phenomenological trilinear
interaction highlighted in this review.  Here we indicated
how the dependence of the exchange interaction on ionic displacements
gives rise to the symmetries expected from Landau theory.  What is
clearly missing is a microscopic calculation of the exchange constants.
This sort of calculation as a function of bond angles
has been pursued for Cu-O-Cu bonds.\cite{SABENA}
However, what is needed here is the more complicated calculation for
Ni-O-Ni bonds and furthermore, it would seem that this is going
to require some sort of calculation based on the local density
approximation to determine the dependence of the exchange tensors on
ionic displacements.  Calculations of this type are being carried out.

\subsection{Closure}

Magnetoelectric materials have been
investigated in depth for the last forty years.  However, we are still
identifying completely new classes of ferroelectric materials, showing
new and different types of magnetoelectric couplings.  With the
increasing interest in materials which couple electric and magnetic
properties, we expect that these magnetoelectric multiferroics will
continue to be an active area of research.

\vspace{0.2 in} \noindent
ACKNOWLEDGEMENTS.  We have greatly profited from many discussions with
A. Aharony, C. Broholm, O. Entin-Wohlman, M. Kenzelmann, T. Kimura,
A. Ramirez, and T. Yildirim.  We acknowledge help from B. Adhikary in
preparing the figures.  ABH thanks the US-Israel BSF for partial support.

\begin{appendix}
\section{QUARTIC TERMS}

\subsection{HTI Phase of NVO}

Since irrep \#4 can not induce any other irrep, the
free energy only involves order parameters of that irrep.  (So
we will omit the superscript $n=4$ which labels that irrep.) Then,
correct to quartic order we write the free energy associated with irrep \#4 as
\begin{eqnarray}
{\cal H} &=& {\cal H}^{(2)} + {\cal H}^{(4)} + \dots \ ,
\end{eqnarray}
where
\begin{eqnarray}
{\cal H}^{(2)} &=& \sum_{\tau \tau'}
v_{\tau ,\tau'} m_\tau (q)^* m_{\tau'} (q)
\end{eqnarray}
and
\begin{eqnarray}
{\cal H}^{(4)} &= (1/4) & \sum_{\tau_1, \tau_2, \tau_3, \tau_4}
w_{\tau_1 ,\tau_2 , \tau_3 , \tau_4} m_{\tau_1} (q)^*
m_{\tau_2} (q)^* m_{\tau_3} (q) m_{\tau_4} (q) \ ,
\end{eqnarray}
where $m_{\tau=1}(q)=m_{xs}(q)$, $m_{\tau=2}(q)=m_{ys}(q)$,
$m_{\tau=3} (q)=m_{zs}(q)$, $m_{\tau=4}(q)=m_{yc} (q)$, and
$m_{\tau=5}(q)=m_{zc}$,  and the spin components at all the sites in the
unit cell are given in Table \ref{NVOSPIN} in terms of these variables.
Hermiticity implies that $v_{\tau, \tau'} = v_{\tau',\tau}^*$ and
$w_{\tau_1 ,\tau_2 , \tau_3 , \tau_4}=w_{\tau_3 ,\tau_4 , \tau_1 , \tau_2}^*$
and $w$ can be taken to be symmetric under interchange of $\tau_1$ and
$\tau_2$ and of $\tau_3$ and $\tau_4$.
As discussed in the text, using inversion symmetry one can show that
all the matrix elements of ${\bf v}$ and ${\bf w}$ are real.

Now we transform to normal modes:
\begin{eqnarray}
m_\tau (q) = \sum_\rho r_{\tau, \rho} \xi_\rho e^{i \phi_\rho} \ ,
\label{MTOXI} \end{eqnarray}
\begin{eqnarray}
\xiv_{\rho'} = \sum_{\tau'} r_{\tau' \rho'} m_{\tau'}(q)
e^{-i \phi_{\rho'}}  \equiv \xi_{\rho'} e^{-i \phi_{\rho'}} \ ,
\label{XITOM} \end{eqnarray}
where $\rho =0,1,2,3,4$ labels the normal mode, the $r$'s are
real, and the critical mode
($\rho_0$) has an amplitude $\xi_0$ which heretofore we called
$\sigma_{\rm HTI}$.  The quartic Hamiltonian is
\begin{eqnarray}
{\cal H}_4 &=& {1 \over 4} \sum_{\tau_1, \tau_2, \tau_3, \tau_4}
\sum_{\rho_1, \rho_2, \rho_3, \rho_4} w_{\tau_1, \tau_2, \tau_3, \tau_4}
r_{\tau_1, \rho_1} r_{\tau_2, \rho_2}
\nonumber \\ && \times
r_{\tau_3, \rho_3} r_{\tau_4, \rho_4}
\xi_{\rho_1} \xi_{\rho_2} \xi_{\rho_3} \xi_{\rho_4} 
e^{i( \phi_{\rho_3} + \phi_{\rho_4} - \phi_{\rho_1} - \phi_{\rho_2})} \ .
\end{eqnarray}
This quartic term will involve contributions proportional to
$\xi_0^p \equiv \sigma_{\rm HTI}^p$, where $p$ ranges from zero to four.
If we were to omit the quartic terms with $p=3$, then the minimum of
the trial free energy would be realized for $\sigma_{\rm HTI} \not= 0$,
but with the other $\xi_\rho$'s being zero.  Therefore, the most important
term to consider is the one cubic in $\sigma_{\rm HTI}$, which is   
\begin{eqnarray}
\delta {\cal H}_4 &=& 
{1 \over 2} \sum_{\tau_1, \tau_2, \tau_3, \tau_4}
\sum_{\rho=1}^3 w_{\tau_1, \tau_2, \tau_3, \tau_4}
r_{\tau_1, 0} r_{\tau_2, 0} r_{\tau_3, 0} r_{\tau_4, \rho} 
\xi_\rho \sigma_{\rm HTI}^3 e^{i (\phi_\rho - \phi_{\rm HTI})} + {\rm c.  c.}
\nonumber \\ &=&
\sum_{\rho=1}^3  A_\rho \xi_\rho \sigma_{\rm HTI}^3 
\cos[\phi_\rho-\phi_{\rm HTI}] \ ,
\end{eqnarray}
where $A_\rho$ is real.  The quadratic terms for the noncritical variables
can be written as
\begin{eqnarray}
\delta {\cal H}_2 &=& {1 \over 2} \sum_{\rho=1}^3
\chi_\rho^{-1} \xiv_\rho^* \xiv_\rho = {1 \over 2}
\sum_{\rho=1}^3 \chi_\rho^{-1} \xi_\rho^2 \ ,
\end{eqnarray}
where $\chi_\rho$ is the susceptibility of the $\rho$th mode.
Then, after minimization with respect to the noncritical variables
we see that $\cos [ \phi_\rho - \phi_{\rm HTI}] = \pm 1$ (so that
$\sigma_{\rm LTI}$ and $\sigma_{\rm LTI}$ are in phase\cite{PHASE}) and
\begin{eqnarray}
\xi_\rho = \chi_\rho A_\rho \sigma_{\rm HTI}^3 \ , \ \ \ \rho > 0 \  .
\end{eqnarray}
Thus the effect of the quartic terms is to induce nonzero
values for the noncritical normal modes and thereby slightly change
the components of the critical eigenvector, but the quartic terms
do not change the fact that all the order parameters $m_\tau$
belong to irrep \#4 and that they all have the same relative phase.

\subsection{LTI Phase of NVO}

Now we consider the LTI phase, where we have two irrep simultaneously present.
There are various types of quartic terms.  First, consider those
quartic terms which only involve a single irrep.  We can apply
the analysis of the HTI phase, to state that such terms do not
modify the conclusion that all the symmetry adapted coordinates
of irrep \#4 have the same phase, $\phi_4$, and all the symmetry
adapted coordinates of irrep \#1 have the same phase $\phi_1$.

Next consider the more general quartic terms which involve both
irreps.  Terms of the type $[m^{(4)}]^* [m^{(1)}]^* m^{(4)} m^{(1)}$
are independent of the phases and therefore after minimization
of the trial free energy these terms do not modify the phases.
There are no terms which involve three order parameters of one
irrep and one order parameter of the other irrep.  So the only
terms which might affect the phases are terms of the form
$[m^{(4)}]^* [m^{(4)}]^* m^{(1)} m^{(1)}$ and its complex conjugate.
So we consider quartic terms of the form
\begin{eqnarray}
F_4 &=& \sum_{\tau_1 \tau_2 \tau_3 \tau_4}
w_{\tau_1 \tau_2 \tau_3 \tau_4}^{(4) (4) (1) (1)}
m_{\tau_1}^{(4)} (q)^* m_{\tau_2}^{(4)} (q)^*
m_{\tau_3}^{(1)} (q) m_{\tau_4}^{(1)} (q) + {\rm c.  c.} \ .
\end{eqnarray}
Hermiticity requires that
$w_{\tau_1 \tau_2 \tau_3 \tau_4}^{(4) (4) (1) (1)}
=[w_{\tau_3 \tau_4 \tau_1 \tau_1}^{(1) (1) (4) (4)}]^*$.
Then inversion symmetry indicates that the $w$ coefficients are real.
Thus these quartic terms give
\begin{eqnarray}
F_4 &=& A \cos[ 2 (\phi_4 - \phi_1)] \sum_{\tau_1 \tau_2 \tau_3 \tau_4}
\sum_{\rho_1 \rho_2 \rho_3 \rho_4}
r_{\tau_1 \rho_1}^{(4)} r_{\tau_2 \rho_2}^{(4)} 
r_{\tau_3 \rho_3}^{(1)} r_{\tau_4 \rho_4}^{(1)}
\xi_{\rho_1}^{(4)} \xi_{\rho_2}^{(4)} 
\xi_{\rho_3}^{(1)} \xi_{\rho_4}^{(1)} \ ,
\end{eqnarray}
where the $r$'s are the real-valued transformation coefficients
determined in quadratic order.  All the quantities in $F_4$ are
real.  So $F_4$ is minimized by either setting $\cos [ 2(\phi_4-\phi_1)] = \pm 1$.  An explicit calculation for the actual experimentally determined
values of the order parameters indicated that the correct choice
of sign is the negative sign, and therefore that the two irreps
are out of phase with one another.  This conclusion agrees with the
intuitive argument based on the idea that quartic terms tend to
enforce the fixed spin length constraint.  When the coordinates of
one irrep are maximal, then those of the other irrep should be minimal.  
Thus we conclude that $|\sin (\phi_{\rm HTI} - \phi_{\rm LTI})|=1$ in
Eq. (\ref{TEMP}).

\section{Do Two Irreps induce a Third One?}

When two irreps, $\Gamma_x(q)$ and $\Gamma_y(q)$ are simultaneously present
(as happens in the LTI phase), one might ask whether their combination could
then induce a third representation, $\Gamma_a(q)$, all  of which are
assumed to be associated with the selected wavevector $q$.  Since
$\Gamma_y(q) \Gamma_y(-q)$ is unity, it is equivalent to ask
whether for some $k$, products like $\Gamma_x(q)^{k-1} \otimes
\Gamma_y(-q)^k \otimes \Gamma_a(q)$
or $\Gamma_y(q)^{k-1} \otimes \Gamma_x(-q)^k \otimes \Gamma_a(q)$
transform like unity.  (The form of this product is dictated by wavevector
conservation.  In this connection we neglect the possible effects of
Umklapp terms.) If one of these products satisfies this condition,
then the existing order parameters can give rise to a linear field acting
on $\Gamma_a(q)$, thereby inducing a nonzero value for this representation.
By explicit enumeration of the various cases one can verify that
the condition to induce a third irrep can not be satisfied.
If, hypothetically, there existed a third phase transition in which a
third irrep condensed, then the presence of these three irreps would
induce the fourth irrep.

\end{appendix}


\begin{thebibliography} {99}
\bibitem{INCOM}
Commensurate systems have a wave vector $q$ such that 
$aq=n/m$, where $n$ and $m$ are integers.  If this condition
is not satisfied, the system is incommensurate.
\bibitem{PHASE}
We say that two complex valued quantities, $z_1$ and $z_2$, have the
{\it same phase} if $z_1/z_2$ is {\it real}, but is not necessarily
{\it positive real}.
\bibitem{ANALOG}
An analogous formulation would be needed for NVO if we were to
consider the effect of the miniscule moments which are induced on
the lower symmetry oxygen sites.
\end{thebibliography}
\end{document}